\documentclass[two column]{aastex63}  
\usepackage{caption}
\usepackage{graphicx}				
\usepackage{amssymb, rotating}
\usepackage{comment}

\begin{document}
\title{Molecular Gas Tracers in Young and Old Protoplanetary Disks}

\author[0000-0002-8310-0554]{Dana E. Anderson}
\affiliation{Earth and Planets Laboratory, Carnegie Institution for Science, 5241 Broad Branch Road, NW, Washington, DC 20015, USA}
\affiliation{Carnegie Postdoctoral Fellow}

\author[0000-0003-2076-8001]{L. Ilsedore Cleeves}
\affiliation{Department of Astronomy, University of Virginia, 530 McCormick Road, Charlottesville, VA 22904, USA}

\author[0000-0003-0787-1610]{Geoffrey A. Blake}
\affiliation{Division of Geological and Planetary Sciences, California Institute of Technology, 1200 E. California Blvd., Pasadena, CA 91125, USA}

\author[0000-0001-8642-1786]{Chunhua Qi}
\affiliation{Harvard-Smithsonian Center for Astrophysics 60 Garden Street Cambridge, MA 02138, USA}

\author[0000-0003-4179-6394]{Edwin A. Bergin}
\affiliation{Department of Astronomy, University of Michigan, 1085 S. University, Ann Arbor, MI 48109, USA}

\author[0000-0003-2251-0602]{John M. Carpenter}
\affiliation{Joint ALMA Observatory, Av. Alonso de C\'{o}rdova 3107, Vitacura, Santiago, Chile}

\author[0000-0002-6429-9457]{Kamber R. Schwarz}
\affiliation{Max-Planck-Institut für Astronomie, Königstuhl 17, 69117 Heidelberg, Germany}

\author[0000-0003-1735-2582]{Claire Thilenius}
\affiliation{Department of Astronomy, University of Virginia, 530 McCormick Road, Charlottesville, VA 22904, USA}

\author[0000-0002-0661-7517]{Ke Zhang}
\affiliation{Department of Astronomy, University of Wisconsin-Madison, 475 N. Charter Street, Madison, WI 53706, USA}

\begin{abstract} Molecular emission is used to investigate both the physical and chemical properties of protoplanetary disks. Therefore, to accurately derive disk properties, we need a thorough understanding of the behavior of the molecular probes we rely on. Here we investigate how the molecular line emission of N$_2$H$\!^+$, HCO$^+$, HCN, and C$^{18}$O compare to other measured quantities in a set of 20 protoplanetary disks. Overall, we find positive correlations between multiple line fluxes and the disk dust mass and radius. We also generally find strong positive correlations between the line fluxes of different molecular species. However, some disks do show noticeable differences in the relative fluxes of N$_2$H$\!^+$, HCO$^+$, HCN, and C$^{18}$O. These differences occur even within a single star-forming region. This results in a potentially large range of different disk masses and chemical compositions for systems of similar age and birth environment. While we make preliminary comparisons of molecular fluxes across different star-forming regions, more complete and uniform samples are needed in the future to search for trends with birth environment or age. \end{abstract}

\section{Introduction}

Extrasolar planetary systems display a wide array of different architectures---evidence for a diverse range of outcomes of the planet formation process \citep{Winn15}. In order to understand the causes of this diversity, we look to the origins of planetary systems and the early stages of their development \citep{Miotello22-arxiv}. Planets form out of the gas and dust which collects in the disk that surrounds a young star during its formation. Similar to the mature planetary systems, these planet-forming or protoplanetary disks are also diverse---differing in size, structure, and composition \citep{Manara22-arxiv}. We seek to characterize these variations in protoplanetary disk properties and evolution to aid comparison with the range of outcomes found in mature planetary systems \citep[e.g.,][]{Mulders21}. But to do this, we first need to develop an understanding of the behavior of the typical bright molecular tracers (e.g.,~CO, N$_2$H$^+$, HCO$^+$, HCN) whose emission we rely on to derive various physical and chemical properties of protoplanetary disks.

Recent surveys of large populations of protoplanetary disks have led to a better understanding of their demographics \citep{Manara22-arxiv} and begun to enable comparisons with the masses of fully evolved exoplanetary systems \citep{Mulders21}. While these surveys focused mainly on observing emission from dust and CO isotopologues, we can further expand our current analysis of protoplanetary disks by utilizing a larger set of distinct molecular gas tracers \citep[see e.g.,][]{Oberg21-maps}. Molecular emission contains information on both the physical and chemical conditions within the disk \citep{Oberg23}, and can be further used to infer disk dynamics \citep[see][]{Henning13}. 

Highly spatially resolved maps of emission from various molecular species reveal unique chemical structures for individual disks \citep{Law21a-maps}. To date such detailed data are available for only a small number of protoplanetary disks. In contrast, surveys of disk-integrated fluxes have been performed over larger numbers of molecular species and disk sources. In combination with physical-chemical modeling of protoplanetary disks, disk-integrated fluxes can be a useful first-look diagnostic of disk gas masses and compositions \citep{Miotello22-arxiv}. But what are the main factors that determine the strength of disk-integrated fluxes? We seek to determine how disk-integrated molecular fluxes relate to other observed disk properties and how they compare among different molecular species. 

Trends have previously been identified in surveys of disk-integrated molecular fluxes focused on C$_2$H, HCN, and isotopologues of CO \citep{Bergner19, Miotello19, Pegues21, Pegues23}. Whereas C$_2$H was found to be correlated with HCN and CN, no correlation was found with CO isotopologue fluxes. In this work, we search for relationships in the disk-integrated fluxes among a suite of molecular species and compare these fluxes with other stellar and disk properties. We compare the disk-integrated molecular fluxes of N$_2$H$^+$, HCO$^+$, HCN, and C$^{18}$O to measured stellar and disk properties for a set of 20 T~Tauri disks. Section~\ref{section:methods} describes our observations and the data taken from the literature.  Section~\ref{section:results} shows the results of our observations and comparisons. We then discuss these results in Section~\ref{section:discussion} and present our conclusions in Section~\ref{section:conc}.

\begin{deluxetable*}{lcccccccccccc}
\tablewidth{0pt}
\setlength{\tabcolsep}{4pt}
\tablecolumns{12}
\tablecaption{\em{Source Properties}\label{table1}}
\tablehead{\multicolumn{1}{l}{Source} & \multicolumn{1}{c}{SpT} & \multicolumn{1}{c}{Dist.} & \multicolumn{1}{c}{Region} &
  \multicolumn{1}{c}{Cont. Freq.} & \multicolumn{1}{c}{Cont. Flux} & \multicolumn{1}{c}{M$_{dust}$$^b$} & \multicolumn{1}{c}{M$_{star}$} & \multicolumn{1}{c}{L$_{star}$} & \multicolumn{1}{c}{Log(M$_{acc}$)} & \multicolumn{1}{c}{R$_{cont,68}$} & \multicolumn{1}{c}{Ref.}  \\
Name$^a$  & & (pc) & & (GHz) & (mJy) & (M$_{\oplus}$) & (M$_{\odot}$) & (L$_{\odot}$) & (M$_{\odot}$) & (\arcsec) & }
\startdata
\multicolumn{5}{l}{Young Star-forming Regions (1--3~Myr)} \\
\hline
 RY Tau & G0 & 141.00 & Taurus & 225.50 & 210.40 & 125.5 & 2.75 & 23.4 & -7.55 & \nodata & [1,2]\\
 UZ Tau Eab$^c$ & M1.9 & 141.00 & Taurus & 225.50 & 129.52 & 77.3 & \nodata & \nodata & \nodata &  0.62 & [1,2] \\
 GG Tau A$^c$ & K7.5 & 141.00 & Taurus & 225.50 & 464.40 & 277.1 & \nodata & \nodata & \nodata & \nodata & [1,2] \\
 IRAS 04302+2247 & \nodata & 161.00 & Taurus & 240.00 & 165.90 & 109.1 & \nodata & \nodata & \nodata & \nodata & [2,3] \\
 DM Tau & M3.0 & 144.05 &  Taurus & 225.50 & 89.4 & 55.7 & 0.291 & 0.241 & -7.992 & 0.92 & [1,2]  \\
 GM Aur & K6.0 & 141.00 & Taurus & 225.50 & 173.20 & 103.3 & 0.69 & 0.903 & -7.97 & 0.87 & [1,2] \\
 Sz 65  & K7 & 153.47 & Lupus & 225.66 & 29.94 & 21.1 & 0.61 & 0.869 & -9.48 & 0.14 & [2,4] \\
 Sz 130 & M2 & 159.18 & Lupus & 225.66 & 1.97 & 1.5 & 0.394 & 0.117 & -9.097 & \nodata & [2,4] \\
 Sz 68$^c$ & K2 & 158.00 & Lupus & 225.66 & 66.38 & 49.6 & 1.32 & 5.691 & -8.12 & \nodata & [2,4] \\
 Sz 90 & K7 & 160.37 & Lupus & 225.66 & 8.74 & 6.7 & 0.73 & 0.419 & -8.93 & 0.1 & [2,4] \\
 J16085324-3914401 & M3 & 163.00 & Lupus & 225.66 & 7.98 & 6.4 & 0.291 & 0.198 & -10.03 & 0.08 & [2,4] \\
 Sz 118 & K5 & 161.46 & Lupus & 225.66 & 23.45 & 18.3 & 0.83 & 0.698 & -9.11 & 0.37 & [2,4] \\
 Sz 129 & K7 & 160.13 & Lupus & 225.66 & 75.90 & 58.3 & 0.73 & 0.421 & -8.27 & 0.31 & [2,4] \\
 IM Lup & K5 & 155.82 & Lupus & 225.66 & 205.00 & 149.1 & 0.72 & 2.514 & -7.85 & \nodata & [2,4] \\
 AS 209 & K5 & 121.25 & Oph N 3a & 239.00 & 288.00 & 108.6 & 0.832 & 1.413 & -7.30 & 1.15 & [5,6,7] \\
 \hline
 \multicolumn{5}{l}{Older Star-forming Regions ($>$5~Myr)} \\
 \hline
 J16142029-1906481 & K5 & 142.00 & Upper Sco. & 340.70 & 40.69 & 8.3 & \nodata & \nodata & \nodata & 0.11 & [2,8] \\
 J16082324-1930009 & K9 & 137.82 & Upper Sco. & 340.70 & 43.19 & 8.3 & 0.646 & 0.319 & -9.138 & 0.25 & [2,8] \\
 J16090075-1908526 & K9 & 137.40 & Upper Sco. & 340.70 & 47.28 & 9.1 & 0.646 & 0.319 & -8.808 & 0.19 & [2,8]\\
 TW Hya & M0.5 & 60.14 & TWA & 233.00 & 558.30 & 55.5 & 0.58 & 0.282 & -8.600 & 0.73 & [6,7,9, \\
 &  &  &  &  &  &  &  &  & &  & 10,11,12] \\
 V4046 Sgr$^c$ & K5+K7 & 71.48 & $\beta$ Pic MG & 235.00 & 338.00 & 46.4 & \nodata & \nodata & \nodata & \nodata & [6,10,13] \\
\enddata
\tablecomments{$^a$The following abbreviations are used in this text: J1614 for J16142029-1906481, J1608 for J16082324-1930009, J1609 for J16090075-1908526, and J16085 for J16085324-3914401. $^b$Calculated in this work using the method from Section~$\ref{section:methods3}$. $^c$Multi-stellar systems.}
\tablereferences{[1] \citet{Herczeg14} for spectral type; [2] \citet{Manara22-arxiv} for all other parameters; [2] \citet{Garufi21} for distance; [3] \cite{vantHoff20} for continuum flux; [4] \citet{Alcala19} for spectral type; [5] \citet{Andrews18_DSHARPI} for all other parameters; [6] \citet{Gaia21} for distance;  [7] \citet{vanderMarel21} for R$_{cont,68}$; [8] \citet{Carpenter14} for spectral type; [9] \citet{Sokal18} for spectral type; [10] \citet{Kastner22-arxiv} for region and V4046~Sgr spectral type; [11] \citet{Tsukagoshi16} for continuum flux; [12] \citet{Venuti19} for M$_{star}$, L$_{star}$, M$_{acc}$; [13] \citet{Kastner18} for continuum flux. }
\end{deluxetable*}

\begin{deluxetable*}{llrrrrrrrrc}
\tablewidth{0pt}
\setlength{\tabcolsep}{4pt}
\tablecolumns{12}
\tablecaption{\em{Disk-integrated Molecular Line Fluxes}\label{table2}}
\tablehead{\multicolumn{1}{l}{}&\multicolumn{1}{l}{Source} 
 & \multicolumn{5}{c}{Line Fluxes  (mJy km s$^{-1}$)$^a$} & \multicolumn{1}{c}{Ref.}  \\
 & Name  & \multicolumn{1}{c}{$^{13}$CO} & \multicolumn{1}{c}{C$^{18}$O} & \multicolumn{1}{c}{HCO$^+$} & \multicolumn{1}{c}{HCN} & \multicolumn{1}{c}{N$_2$H$^+$} }
\startdata
1 & J16142029-1906481 & 27.7$\pm$3.0(1) & 7.5$\pm$1.1(1) & 78.2$\pm$8.1(1) & $<$8.4(1) & $<$5.8(1)  & [1,2,3] \\
   & & 7.5$\pm$2.2(1)$^{*}$ & & & & & \\
2 & J16082324-1930009 & 30.7$\pm$3.2(1) & 7.9$\pm$1.5(1) & 62.8$\pm$6.4(1) & 10.43$\pm$1.07(2) & 18.9$\pm$2.1(1) & [1,2,3,4] \\
 & & 11.8$\pm$2.6(1)$^{*}$ & & & & & \\
3 & J16090075-1908526 & 39.9$\pm$4.1(1) & 13.6$\pm$1.8(1) & 57.0$\pm$5.8(1) & 59.6$\pm$6.2(1) & 10.2$\pm$1.5(1) & [1,2,4]\\
4 & RY Tau & 15.35$\pm$2.89(2)$^{*}$ & $<$7.31(2)$^{*}$ & 13.58$\pm$2.64(2) & 7.34$\pm$1.76(2) & $<$5.45(2) & [1,5]\\
5 & UZ Tau Eab & 8.98$\pm$2.16(2)$^{*}$ & $<$5.41(2)$^{*}$ & 7.73$\pm$1.99(2) & $<$5.38(2) & $<$5.80(2) & [1,5] \\
6 & GG Tau A & 43.66$\pm$5.20(2)$^{*}$ & 8.08$\pm$2.11(2)$^{*}$ & 46.49$\pm$5.33(2) & 24.45$\pm$3.09(2) & 10.60$\pm$2.40(2) & [1,5] \\
7 & IRAS 04302+2247 & 16.74$\pm$1.75(3)$^{*}$ & 37.31$\pm$4.90(2)$^{*}$ & 14.28$\pm$1.45(3) & 23.12$\pm$2.80(2) & 17.61$\pm$2.20(2) & [1,5] \\
8 & Sz 65  & 9.71$\pm$1.61(2) & 4.15$\pm$1.13(2) & 21.9$\pm$2.3(1) & $<$3.2(1) & $<$1.9(1) & [6,7,8] \\
 & & $<$1.03(2)$^{*}$ & $<$6.0(1)$^{*}$ & & & & \\
9 & Sz 130 & 47.0$\pm$8.5(1) & $<$1.18(2) & 18.4$\pm$2.0(1) & $<$3.7(1) & $<$2.7(1) & [6,7,8] \\
  & & $<$9.6(1)$^{*}$ & $<$5.1(1)$^{*}$ & & & & \\
10 & Sz 68 & 9.15$\pm$1.61(2) & 4.44$\pm$1.39(2) & 18.3$\pm$2.0(1) & $<$2.4(1) & $<$1.5(1) & [6,7,8] \\
 & & $<$1.21(2)$^{*}$ & $<$6.9(1)$^{*}$ & & & & \\
11 & Sz 90 & 4.33$\pm$1.07(2) & $<$1.84(2) & 29.8$\pm$3.1(1) & 16.9$\pm$2.6(1) & 7.0$\pm$1.0(1) & [6,7,8] \\
 & & $<$7.8(1)$^{*}$ & $<$5.4(1)$^{*}$ & & & & \\
12 & J16085324-3914401 & 26.7$\pm$5.3(1) & $<$1.36(2) & 25.4$\pm$2.6(1) & 16.2$\pm$1.9(1) & $<$3.6(1) & [6,7,8] \\
  & & $<$7.5(1)$^{*}$ & $<$5.7(1)$^{*}$ & & & & \\
13 & Sz 118 & 6.95$\pm$1.50(2) & $<$2.36(2) & 37.6$\pm$3.8(1) & 17.5$\pm$2.1(1) & 6.4$\pm$1.1(1) & [6,7,8] \\
  & & 8.0$\pm$2.7(1)$^{*}$ & $<$5.1(1)$^{*}$ & & & & \\
14 & Sz 129 & 51.6$\pm$8.9(1) & $<$1.09(2) & 59.5$\pm$6.1(1) & 44.3$\pm$4.6(1) & 20.3$\pm$2.7(1) & [6,7,8] \\
 & & $<$6.3(1)$^{*}$ & $<$4.8(1)$^{*}$ & & & & \\
15 & IM Lup & 11.5$\pm$1.7(3) & 27$\pm$4(2) & 9.67$\pm$1.01(3) & 56.54$\pm$5.70(2) & 20.40$\pm$2.10(2) & [9,10,11,12,13] \\
 & & 83.70$\pm$8.41(2)$^{*}$ & 15.92$\pm$1.70(2)$^{*}$ & & & & \\
16 & DM Tau & & 30$\pm$3(2) & 52.3$\pm$5.4(2) & 29.66$\pm$4.00(2) & 12.86$\pm$1.35(2) & [11,13,14,15,16]  \\
 & & 6.84$\pm$1.04(3)$^{*}$ & 11.16$\pm$1.10(2)$^{*}$ & & & & \\
17 & GM Aur & 50.28$\pm$5.05(2)$^{*}$ & 10.92$\pm$1.16(2)$^{*}$ & 43.7$\pm$4.7(2) & 18.35$\pm$2.12(2) & 14.87$\pm$1.50(2) & [10,11,16] \\
18 & AS 209 & 22.69$\pm$2.30(2)$^{*}$ & 53.8$\pm$6.0(1)$^{*}$ & 40.2$\pm$4.3(2) & 37.07$\pm$3.80(2) & 91.7$\pm$9.3(1) & [10,11,12,13] \\
19 & TW Hya & 46.1$\pm$4.6(2) & 12.1$\pm$1.2(2) & 12.9$\pm$2.1(3) & 8.5$\pm$1.7(3) & 20.0$\pm$2.1(2) & [15,17,18] \\
 & & 27.6$\pm$1.8(2)$^{*}$ & 57$\pm$6(1)$^{*}$ & & & & \\
20 & V4046 Sgr & 84.4$\pm$9.6(2)$^{*}$ & 12.03$\pm$1.20(2)$^{*}$ & 11.43$\pm$1.17(3) & 10.000$\pm$1.000(3) & 37.61$\pm$3.85(2) & [11,12,13,19] \\
\enddata
\tablecomments{$^a$In this table, flux values (i$\pm$j)$\times$10$^{k}$ are abbreviated as i$\pm$j(k). Fluxes are reported for the $J$~=~3--2 transition, except for $^{13}$CO and C$^{18}$O $J$~=~2--1 fluxes indicated with an asterisk (*). Line flux values are listed here at the source distance but scaled to a common distance of 160~pc prior to further comparison analyses.}
\tablereferences{[1] This work, as described in Section~\ref{section:methods2}; [2] ALMA 2018.1.01623.S; [3] \citet{Bergner20}; [4] ALMA 2015.1.01199.S; [5] SMA 2018B-S046;  [6] \citet{Ansdell16}; [7] \citet{Ansdell18}; [8] \citet{Anderson22}; [9] \citet{Cleeves16}; [10] \citet{Oberg21-maps}; [11] \citet{Qi19}; [12] \citet{Oberg11}; [13] \citet{Bergner19}; [14] \citet{Guilloteau13}; [15] \citet{Zhang19}; [16] \citet{Guilloteau16}; [17] \citet{Cleeves15}; [18] \citet{Trapman22}; [19] \citet{Kastner18} }
\end{deluxetable*}

 \begin{figure*}
 \includegraphics[width=\linewidth]{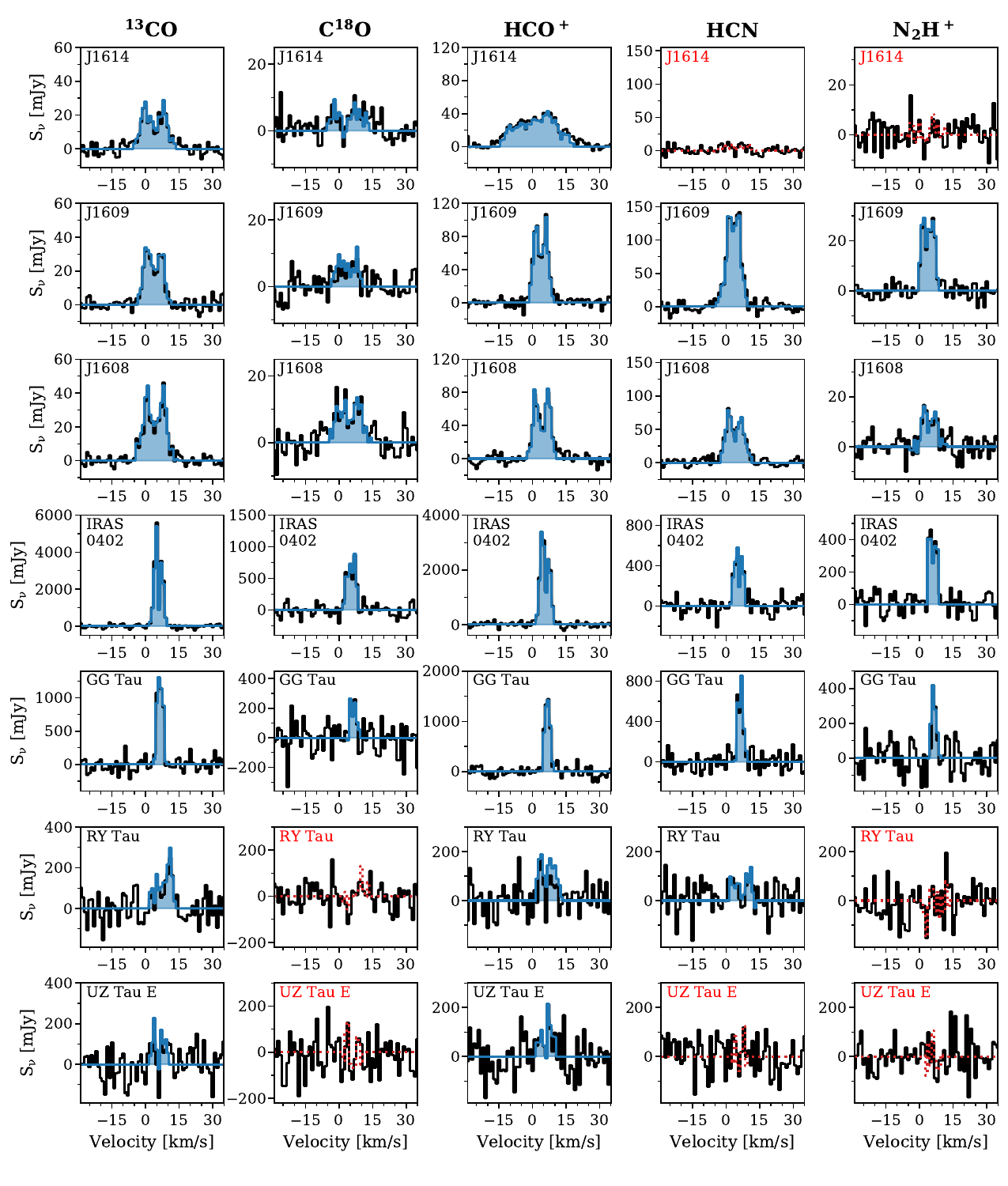}
 \caption{Spectra showing $^{13}$CO, C$^{18}$O ($J$~=~3--2 for the top three rows, 2--1 for the bottom four), HCO$^+$, HCN, and N$_2$H$^+$ ($J$~=~3--2) line emission from the observed disks described in Sections~\ref{section:methods1a}--\ref{section:methods1b}. Black curves show the integrated flux from within the channel-summed empirical mask, whereas the integrated fluxes within the channel-by-channel empirical masks are shown in blue (line detections) or red (non-detections). Empirical masks are based on bright line emission as described in Section~\ref{section:methods2}.} \label{fig:spectra}
 \end{figure*}

\section{Observations}\label{section:methods}
Our disk sample represents 20 T~Tauri stars from nearby star-forming regions (see Table~\ref{table1}). This includes 7 sources with new Atacama Large Millimeter/submillimeter Array (ALMA) and Submillimeter Array (SMA) observations of molecular line fluxes as described in Sections~\ref{section:methods1a}--\ref{section:methods2}. The remaining data are taken from the literature as described in Section~\ref{section:methods3}. This mixed sample includes sources from younger (1--3~Myr-old; Lupus, Taurus, and Oph N 3a) and older ($>$5~Myr-old; Upper Scorpius, TWA, and the $\beta$ Pic Moving Group [MG]) star-forming regions with observations of HCO$^+$, HCN, and N$_2$H$^+$ $J$~=~3--2 and $^{13}$CO and C$^{18}$O $J$~=~3--2 or 2--1 spectral lines. We incorporate sources from the literature based on the availability of published line fluxes for these select molecular species and transitions at the time of this analysis. Future studies may augment this sample through thorough mining of archival data. \footnote{We exclude AS~205 due to the lack of agreement among the literature on the star-formation region with which it is associated \citep[e.g.,][]{2014ApJ...792...68S,Barenfeld16,Andrews18_DSHARPI,Kurtovic18_DSHARPIV}.}

\subsection{ALMA Observations}\label{section:methods1a}
Three sources with detected CO emission and relatively high dust masses were selected from the ALMA survey of Upper Sco by \cite{Barenfeld16}. N$_2$H$^+$ observations of J1608 and J1609 were taken in 2016 (ALMA Cycle~3, PI:Anderson, 2015.1.01199.S) and previously reported by \cite{Anderson19}. A third source, J1614, was chosen to represent a disk with a similar dust mass but higher CO flux (by a factor of 5--20$\times$) than the other two. In addition to searching for N$_2$H$^+$ in J1614, we observed HCO$^+$, HCN, $^{13}$CO, and C$^{18}$O $J$~=~3--2 line emission from all three sources during ALMA Cycle~6 (PI:Anderson, 2018.1.01623.S). 

Band~6 observations of HCO$^+$ and HCN were taken on December 4, 2018. Band~7 observations of N$_2$H$^+$ were taken on December 8, 2018. Band~7 observations of $^{13}$CO and C$^{18}$O were taken on March 8 and 14--15, 2019. Observations were taken with 43--47 antennas and maximum baselines of 740.4, 783.5, and 313.7--360.6~m, respectively. Spectral setups included multiple windows to capture line and continuum emission as shown in Table~\ref{tableA1}. Observed channel widths were 244.141~kHz ($\sim$0.25~km~s$^{-1}$) for spectral line windows and 976.562 or 15625~kHz for continuum windows. The on-source integration time per source was about 23~min for Band~6, 36~min for Band~7 observations of N$_2$H$^+$ and 57~min for Band~7 observations of $^{13}$CO and C$^{18}$O. 

The data were calibrated using the standard ALMA pipeline and analyzed with the Common Astronomy Software Applications (CASA) package \citep{2022PASP..134k4501C}. When necessary, calibrated measurement sets were restored using the version of CASA specified in the dataset documentation available on the ALMA archive. Otherwise, imaging and analysis were performed using CASA version 6.4.0. Pipeline flux and bandpass calibrators were J1427-4206 in Band~6 and J1517-2422 in Band~7. J1625-2527 was used as the pipeline phase calibrator. Self-calibration was tested but ultimately not applied to the data because it did not produce improvements in signal-to-noise that were significant for this analysis. 

\subsection{SMA Observations}\label{section:methods1b}
Four sources with strong detections of $^{13}$CO and HCO$^+$ emission were selected from the IRAM~30-m survey of Taurus by \cite{Guilloteau13,Guilloteau16}. These sources represent a younger age group relative to Upper Sco, including the Class~I source IRAS~04302+2247 \citep{Garufi21}, and included binary/multi systems (see Table~\ref{table1}) to add more diversity to the total sample. Using the Submillimeter Array (PI:Anderson, 2018B-S046), we searched for N$_2$H$^+$, HCO$^+$, and HCN $J$~=~3--2 and CO, $^{13}$CO, and C$^{18}$O $J$~=~2--1 line emission from our sources. Data were taken between December 16--19, 2018 with 28 baselines (15 baselines on Dec.~19) in the ``compact" array configuration, providing an angular resolution of 3--4\arcsec. Each source was observed for a total of 5--6~hr. The LO frequencies were 272.489~GHz for RxA receiver (345~GHz setting) and 225.119~GHz for RxB receiver (240~GHz setting), covering our selected suite of molecular lines and several additional species including C$_2$H (not presented here). 

The SWARM correlator provides 8~GHz bandwidth per sideband, divided into four equal sized chunks with uniform spectral resolution of 140~kHz (0.18~km~s$^{-1}$ at 230~GHz and 0.15~km~s$^{-1}$ at 280~GHz). Calibration of visibility phases and
amplitudes was performed using periodic observations of quasars 3C111 and 0510+180, at 15 minute intervals. Measurements of Uranus were used to obtain an absolute scale for calibration of the flux densities. 3C279 was used as the bandpass calibrator. All data were phase- and amplitude-calibrated using the MIR software package \footnote{\url{https://www.cfa.harvard.edu/~cqi/mircook.html}}. The calibrated SMA data were then exported in uvfits format for subsequent imaging and analysis using the CASA software package version 6.4.0. The following data were flagged:  268.0--268.5~GHz in the lower sideband and 276.5--277.0~GHz in the upper sideband for Antenna~3 in the 345~GHz receiver data and 220.8--221.2~GHz in the lower sideband and 229.1--229.5~GHz in the upper sideband for Antenna~2 in the 240~GHz receiver data.

\subsection{Spectral Line Analysis}\label{section:methods2}

Line fluxes were derived from the new observations described in Sections~\ref{section:methods1a}--\ref{section:methods1b} using the following steps. In addition, N$_2$H$^+$ line fluxes were re-derived from the 2015.1.01199.S dataset to produce values using the same methodology. Continuum subtraction was performed on the spectral line windows using CASA's $uvcontsub$. For each source and molecular line, an initial empirical mask was generated based on a smoothed version of the ``dirty" image cube (an image cube generated with zero iterations of the CLEAN algorithm). The dirty image cube was smoothed via convolution with a Gaussian beam with dimensions 1.5$\times$ the original beam size. For each channel in the smoothed dirty image cube, we selected pixels with flux densities $>$3$\times$ the standard deviation of a sample of off-source pixels. For the SMA data, the same procedure was used but without smoothing. The largest 2D clump of selected pixels in each channel is taken as the initial mask shape for that channel. We exclude any clumps in spatial regions (or channels) that are too far separated from the central source region (or velocity) to be probable line emission based on visual inspection of the images. For the relatively weak C$^{18}$O emission (and N$_2$H$^+$ emission in J1608) where empirically-derived masks were overly influenced by noise, the mask generated from the stronger $^{13}$CO emission was used. The $^{13}$CO mask was also used in determining the upper limits on fluxes for molecular lines that were not detected. For GG~Tau~A, RY~Tau, and UZ~Tau~E, empirical masks generated from the strongest emission line, CO $J$~=~2--1 (see appendix Figure~\ref{fig:CO_spectra}) were used for all other lines.

Images are generated by CASA's $tclean$ using Briggs weighting with a robust parameter of 0.5, a threshold of roughly 3$\times$ the standard deviation of the flux density in a sample of line-free channels, and channel widths of 1~km~s$^{-1}$. We provided an input mask to $tclean$ based on the empirical mask generated for each line as described above. For most lines, we used the empirical mask described above as the input mask. For the weak and non-detected lines where the input mask used was that of a brighter line, a channel-summed version of the bright line mask is used. This is done as a precaution to avoid introducing a fake line signature via the $tclean$ mask. To generate the channel-summed mask, the initial empirical bright line mask components in each channel are summed over all channels. This summed mask is applied over the range of channels where line emission is observed. The line emission velocity ranges are -5--15~km~s$^{-1}$ for Upper Sco sources (-15--20~km~s$^{-1}$ for HCO$^+$ in J1614 given its unusually wide emission line as shown in Figure~\ref{fig:spectra}), 4--9~km~s$^{-1}$ for GG~Tau~A, 2--8~km~s$^{-1}$ for IRAS~04302+2247, 2--10~km~s$^{-1}$ for RY~Tau, and 2--13~km~s$^{-1}$ for UZ~Tau~E. Listed velocity ranges are inclusive for 1~km~s$^{-1}$ channels.

Spectra are generated by multiplying the final image by a specific mask using CASA's $immath$ and producing a spectrum of the product using $specflux$. For the sufficiently bright and largely unresolved line emission in this work, the empirical masks based on the dirty images do not significantly differ from those derived from the clean images when using the same approach and remain appropriate for this step. The black lines in Figure~\ref{fig:spectra} show spectra generated using a channel-summed version of each line's empirical mask over the entire image cube. The blue solid (for detections) and red dotted (for non-detections) lines show the flux collected in each individual channel of the original (non-channel-summed empirical mask. The blue spectra are integrated over the line velocity range to produce the flux estimates given in Table~\ref{table2}. Noise estimates are determined by applying each line's empirical mask to 30 different line-free regions across the image cube. Because some of the selected line-free regions are off the phase center, the primary beam correction using CASA's $impbcor$ was applied to the image before the noise and line flux estimates were made. The spatial regions sampled are selected to be as close to the source region as possible without sampling line emission to mitigate overestimation of the noise, which is dependent on the distance from the phase center. In our case, the primary beam correction had almost no effect on the line flux estimates. The standard deviation of fluxes from the 30 line-free regions is added in quadrature with 10\% of the integrated line flux representing the assumed uncertainty for the flux amplitude calibration \citep{cortes_paulo_2023_7822943} and provided as the uncertainties listed in Table~\ref{table2}. Upper limits are estimated as three times this uncertainty value with fluxes above this threshold being considered detections.

\subsection{Data from the Literature}\label{section:methods3}

Our analysis combines the observations described above with data from the literature. Here we describe the literature data that has been included. Available source distances are from \cite{Manara22-arxiv}, which are derived by inverting their parallaxes from Gaia Data Release~3 \citep{Gaia21}. Values for AS~209, TW~Hya, and V4046~Sgr (which are not included in \citealt{Manara22-arxiv}) are derived in the same manner. For IRAS~04302+2247, the distance is taken to be 161~pc \citep{Garufi21}. Most sources belong to the Lupus ($\lesssim$3~Myr), Taurus ($\sim$1--3~Myr), and Upper Scorpius ($\sim$5--10~Myr) star-forming regions \citep{Manara22-arxiv}. AS~209 resides to the northeast of the main Ophiuchus region in Oph~N~3a, with an estimated age of $\sim$1 (0.4--2.5)~Myr \citep{Andrews18_DSHARPI}. TW~Hya and V4046~Sgr are located in the nearby TW~Hydra Association ($\sim$8~Myr) and $\beta$~Pic Moving Group ($\sim$24~Myr), respectively \citep{Kastner22-arxiv}. 

\begin{figure*}
 \includegraphics[width=\linewidth]{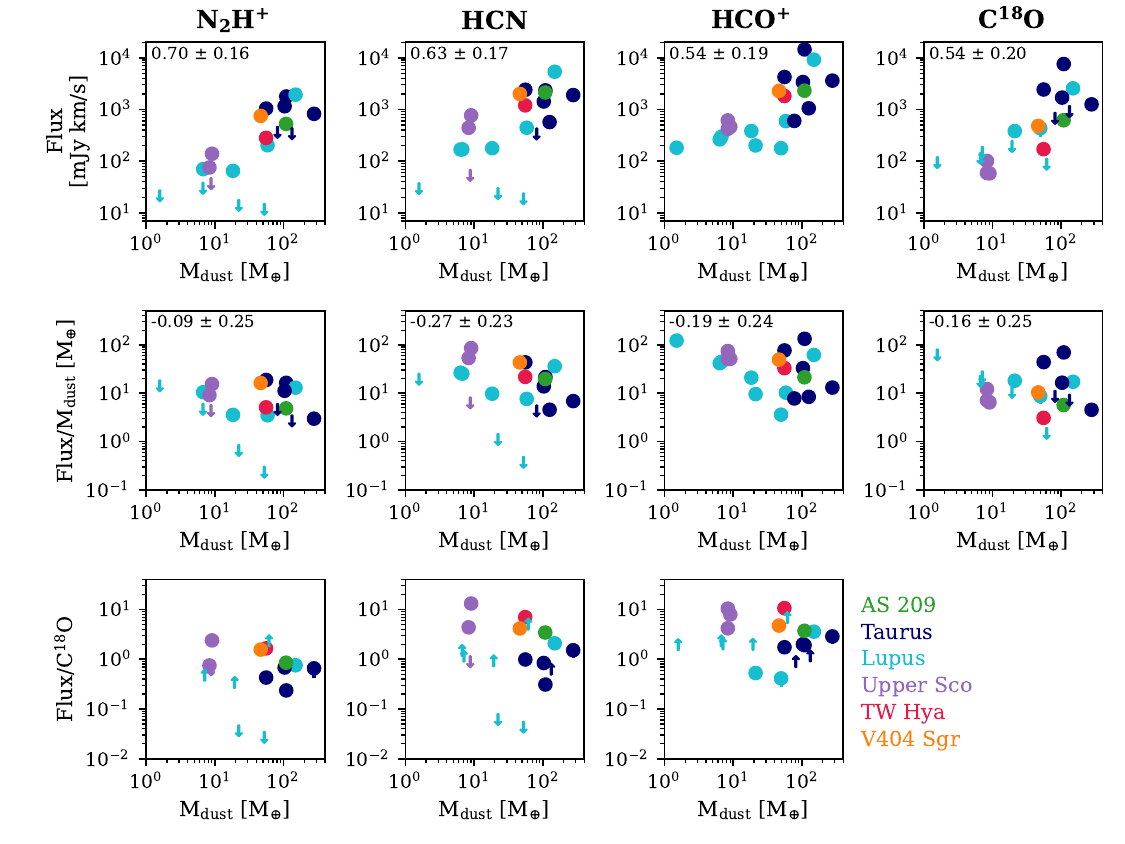}
 \caption{N$_2$H$^+$, HCN, HCO$^+$, and C$^{18}$O $J$~=~3--2 observed line fluxes (columns) compared to disk dust masses (first row). Line fluxes are also normalized by the dust mass (second row) or C$^{18}$O line flux (third row). Fluxes are scaled to a distance of 160~pc. Arrows represent upper or lower limits. Colors indicate the source or star-forming region as shown by the key in the bottom right. Computed correlation coefficients are provided in the upper left of each panel. Correlation coefficients are not provided for the final row because there are no upper or lower bounds on flux ratios. See discussion of this figure in Section~\ref{section:results2}.} 
 \label{fig:FluxvsMdust}
 \end{figure*}

Dust masses (Table~\ref{table1}) are computed as described in \cite{Manara22-arxiv} with the formula used by \cite{Ansdell16} from \cite{Hildebrand83}: 
\begin{displaymath} M_{dust}=\frac{F_{\nu}d^2}{\kappa_{\nu}B_{\nu}(T_{dust})}\end{displaymath}
where $\kappa_{\nu}$~=~2.3($\nu$/230~GHz) cm$^{2}$ g$^{-1}$ and assuming a dust temperature of 20~K. This assumes optically-thin, isothermal dust emission at (sub-)millimeter wavelengths. The sub-mm dust emission may in fact be optically thick \citep[e.g.,][]{Zhu19,Sierra21,Xin23}, resulting in higher dust masses and complicating our search for trends in this analysis since we would not be tracking the true dust mass. Continuum fluxes from \cite{Manara22-arxiv} are used when available. In addition, the continuum flux data from the following works are included: \cite{Andrews18_DSHARPI} for AS~209, \cite{vantHoff20} for IRAS~04302+2247, \cite{Tsukagoshi16} for TW~Hya, and \cite{Kastner18} for V4046~Sgr. The selected continuum fluxes used here are those measured at frequencies of 225--240~GHz, with the exception of Upper Sco for which the continuum fluxes were measured at 340.7~GHz. The uncertainty in dust mass (see error bars in Figure~\ref{fig:FluxvsMdust}) is taken as a 10\% systematic uncertainty in flux amplitude calibration \citep[e.g.,][]{Andrews13, Andrews18_DSHARPI} converted into a mass value using the formula above. 

The ALMA observations of GG~Tau~A by \cite{Akeson19} result in a dust mass of $\sim$4~M$_{\oplus}$ for the circumstellar disk around GG~Tau~Aa. At their high angular resolution (with a beam of 0.18\arcsec$\times$0.16\arcsec), they are able to separate the circumstellar emission from the wider ring around the GG~Tau~A multiple system. Given that our line flux observations of GG~Tau~A were taken at a lower angular resolution and we do not separate out these different components, we use the larger continuum flux value measured by \cite{Andrews13} for the dust mass ($\sim$277~M$_{\oplus}$) in our analyses. 

Spectral types, stellar luminosities, stellar masses, mass accretion rates, and mm dust disk sizes (based on the radius enclosing 68\% of the total continuum flux, R$_{cont,68}$) are taken from \citet[][and references therein]{Manara22-arxiv} when available. See Table~\ref{table1} for additional references used for data on AS~209, TW~Hya, and V4046~Sgr. Also note that Sz~68, GG~Tau~A, UZ~Tau~E, and V4046~Sgr are all binary or multiple-star systems, which may complicate the interpretation of their data and behavior relative to single stars.

Line fluxes are taken from the literature referenced in Table~\ref{table2}. In cases where multiple literature values were available, higher sensitivity observations with stronger signals in the observed spectra are preferred. When multiple analyses were performed on the same or similar datasets, the value derived using a Keplerian mask approach is used. Most fluxes derived from datasets with similar sensitivities are consistent across different methods used in the literature, but a few varied by up to $\sim$2$\times$ (e.g., HCN in IM~Lup, C$^{18}$O $J$~=~2--1 in V4046~Sgr). Uncertainty values listed in Table~\ref{table2} include flux calibration uncertainties of 10--15\% as specified in the literature referenced in the final column of the table (this systematic uncertainty was added in quadrature to previously reported uncertainties if not already included). 

Fluxes shown in the analyses in this work (Section~\ref{section:results}) are for the $J$~=~3--2 transition for each molecule. When only $^{13}$CO and C$^{18}$O $J$~=~2--1 fluxes are available, they are multiplied by a factor of 2$\times$ as a conversion to approximate the $J$~=~3--2 fluxes. The accurate value for this conversion is uncertain. For optically-thin emission in local thermodynamic equilibrium (LTE), this conversion would be $\sim$4--5$\times$ for 20--30~K gas. We base our value on the observed ratios of C$^{18}$O 3--2 / 2--1 that are available for three sources in our sample. These values are in the range of $\sim$1--3, not well described by an optically-thin LTE analysis. Line fluxes in the following analyses (Section~\ref{section:results}) are corrected for differences in source distances by scaling to a uniform distance of 160~pc.  

 \begin{figure*}
 \includegraphics[width=\linewidth]{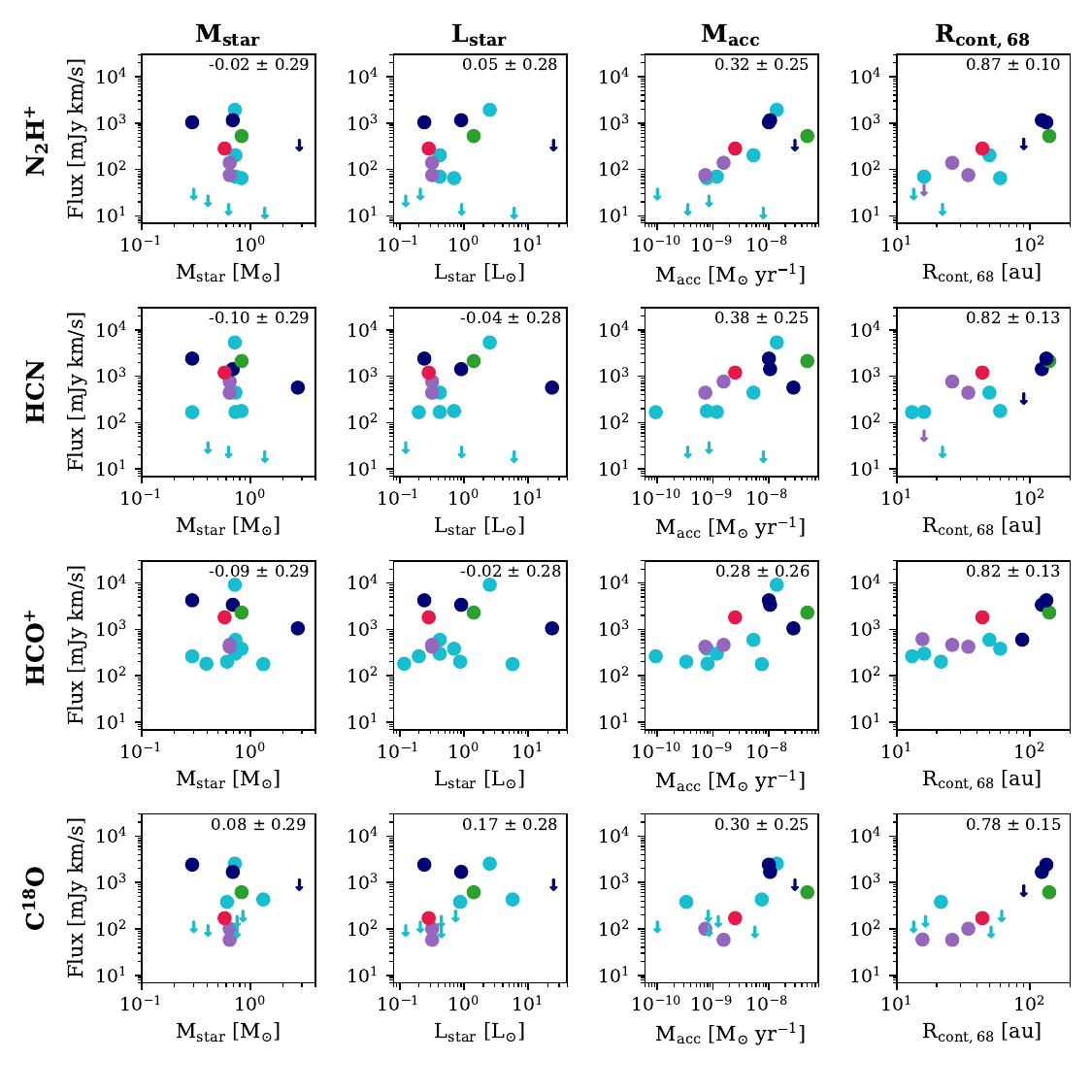}
 \caption{N$_2$H$^+$, HCN, HCO$^+$, and C$^{18}$O $J$~=~3--2 observed line fluxes (rows) compared to stellar masses (first column), stellar luminosities (second column), mass accretion rates (third column), and outer disk radii seen in (sub-)millimeter dust (fourth column). Fluxes are scaled to a distance of 160~pc. Arrows represent upper limits. Colors indicate the source or star-forming region as indicated in Figure~\ref{fig:FluxvsMdust}. Only a subset of the sample is shown per panel as literature data were not equally available for all sources (see Table~\ref{table1}). Computed correlation coefficients are provided in the upper right of each panel. See discussion of this figure in Section~\ref{section:results3}.}
 \label{fig:FluxvsSourceProp}
 \end{figure*}

\section{Results}\label{section:results}
Using the newly measured line fluxes in combination with preexisting data from the literature, we search for any relationships among physical and chemical source properties. Our sample includes a maximum of 20 sources per comparison (based on availability of data, see Table~\ref{table1}) across different star-forming regions. Five sources---three from Upper Sco, TW Hya, and V4046~Sgr---are $>$5~Myr old, representing a later stage of disk evolution relative to the 15 younger disks (1--3~Myr old) in our sample.

\subsection{Line Detections from New Observations}
First, we present the disk integrated line fluxes measured using the method described in Section~\ref{section:methods2} for sources 1--7 in Table~\ref{table2}. Figures~\ref{fig:spectra} and \ref{fig:maps} show the spectra and velocity-integrated line intensity (moment 0) maps, respectively, for all seven sources. We detected N$_2$H$^+$ in IRAS~04302+2247 and GG~Tau~A, in addition to the previously identified N$_2$H$^+$ emission in two of the Upper Sco sources \citep{Anderson19}. HCO$^+$ is detected in all seven sources, while HCN is detected in five. $^{13}$CO and C$^{18}$O are both detected in all three Upper Sco disks. $^{13}$CO is detected in all four of the Taurus disks, but C$^{18}$O is only clearly detected in two. Note that the differences in detection rates between the Upper Sco and Taurus samples are due to the difference in requested sensitivity of the observations for each survey (see Sections~\ref{section:methods1a}--\ref{section:methods1b}). For the Taurus disks, the main isotopologue of CO is also detected in all four disks (see Figure~\ref{fig:CO_spectra}). Additional molecules detected but not discussed in this work are CN in all three Upper Sco disks and C$_2$H and H$_2$CO in GG~Tau~A.

\subsection{Searching for Line Flux Trends}
With the combined list of newly measured disk-integrated line fluxes and those from the literature, we search for any relationships between line fluxes (or flux ratios) and other observed stellar and disk properties. The references for stellar and disk properties used in this analysis are provided in Table~\ref{table1} and Section~\ref{section:methods3}. To provide an appropriate comparison of flux values, fluxes are scaled to a common distance of 160~pc for these analyses. Fluxes are multiplied by $\frac{d^2}{(160~pc)^2}$ where $d$ is the source distance in parsecs from Table~\ref{table1}. Using the Python linear regression tool \texttt{linmix}\footnote{https://github.com/jmeyers314/linmix} based on the method of \cite{Kelly07}, we test for linear correlations between pairs of parameters. \cite{Kelly07} produced a Bayesian method that takes into account measurement errors in two variables and nondetections in one variable in linear regression. We apply this method to investigate trends with fluxes and normalized fluxes where we have nondetections in one variable. The method uses Markov chain Monte Carlo (MCMC) sampling to approximate the posterior distribution. The computed linear correlation coefficient ranges from -1 for a strong negative correlation to 1 for a strong positive correlation, with a value of zero indicating no correlation between the two variables. We report the median correlation coefficient from the \texttt{linmix} runs with one standard deviation as the listed uncertainties. A summary of the computed correlation coefficients is provided in Figure~\ref{fig:stats}.

\subsubsection{Line Fluxes vs.~Dust Mass}\label{section:results2}
The (sub-)millimeter dust mass provides an indication of the amount of solid material present in the disk. Meanwhile, the four molecular tracers track different components of the disk gas. By comparing these line fluxes with the disk dust mass we can investigate whether these various gas components and the dust are related or independent. Note that these trends may be affected by the optical thickness of the dust and/or line emission. $^{13}$CO and HCO$^+$ are expected to be the most abundant of the selected species and the most likely to be optically thick, although in sufficiently bright sources emission from other molecular species may be optically thick as well. \cite{Bergner19} find HCN $J$~=~3--2 emission to be optically thick in AS~209 and V4046~Sgr based on the H$^{13}$CN/HCN ratio. Here we choose to make comparisons with C$^{18}$O rather than $^{13}$CO because while both trace the CO abundance, the rarer isotopologue C$^{18}$O is expected to be less effected by optical depth effects. \cite{Anderson22} found no trend between line fluxes and the dust mass of seven disks in Lupus (numbered 8--14 in Table~\ref{table2}). In the present larger sample, the top row of Figure~\ref{fig:FluxvsMdust} shows that the line fluxes of all four molecular species generally increase with the disk dust mass. This is consistent with the idea that disks with more dust have more material overall and therefore stronger molecular emission. 

 There appears to be a change to a steeper slope in line flux vs.~dust mass for disks with more than $\sim$10--100~M$_{\oplus}$ mass in dust. While this is particularly apparent in the HCO$^+$ emission, whether the same trend exists for N$_2$H$^+$, HCN, and C$^{18}$O is ambiguous given the current nondetections. The canonical mass of solid material needed to initiate runaway gas accretion leading to giant planet formation is $\sim$10~M$_{\oplus}$ \citep[although it depends on disk parameters as shown by][]{Piso14}. This break in the line emission vs.~dust mass could indicate a difference in the gas-to-dust mass ratio between systems that currently have enough solid material to undergo future gas giant planet formation vs.~those that do not. Disks with $<$10~M$_{\oplus}$ of solids may have already undergone giant planet formation or they may have never contained sufficient solid material to allow for giant planet formation from their onset. The apparent change in slope could alternatively be the result of the transition from optically thin to optically thick emission, where line fluxes transition from tracking the column density of molecular gas to the gas temperature. As a the molecular gas column of a species increases, it increases the likelihood of its emission becoming optically thick.  We expect this transition to occur within the set of observed fluxes, but deriving the exact location for each molecular transition would require more knowledge of the individual physical disk structures. In addition, assuming the correlation between the molecular gas column densities and M$_{dust}$ is stronger than the correlation between the emitting layer gas temperatures and M$_{dust}$, one would expect the optically thin emission to have a stronger positive correlation with M$_{dust}$ than the optically thick emission. The trend seen in Fig.~\ref{fig:FluxvsMdust} defies this expectation.

For N$_2$H$^+$, HCN, and C$^{18}$O, the significance of the increasing trends in molecular line fluxes with M$_{dust}$ are affected by several upper limits in the measured line fluxes. The median correlation coefficients indicate a tentative to moderately significant positive linear correlation between the molecular fluxes and M$_{dust}$ for all four molecular species. Computed correlation coefficients are listed in the upper left-hand corner of the individual panels in Fig.~\ref{fig:FluxvsMdust}.

When molecular fluxes are normalized by M$_{dust}$ (Fig.~\ref{fig:FluxvsMdust}, row 2), they show a relatively flat trend with two or more orders of magnitude in scatter relative to M$_{dust}$. The flat trend suggests that within the limits of this sample there are no differences in the M$_{dust}$-normalized line flux ratios between disks with smaller vs.~larger dust masses. Worthy of note, there is no clear change in the normalized line flux indicating where the transition from optically thin to optically thick line emission might occur. Moreover, the large scatter in the dust-normalized line fluxes suggests that while the line fluxes have some dependence on M$_{dust}$, there are still other factors that control these molecular line fluxes. 

To search for potential variations in the chemical composition of the disk gas vs.~M$_{dust}$, we examine the line flux ratios of N$_2$H$^+$, HCN, and HCO$^+$ over the C$^{18}$O line flux. Sources are excluded from the plot when neither species in the ratio is detected. Here we generally find a flat trend with scatter ranging more than two orders of magnitude---revealing no clear distinctions in gas-phase chemical abundances as a function of changing dust mass (as indicated by molecular line fluxes), but rather a large amount of variation across the whole dust mass range. 

\begin{figure*}
\includegraphics[width=\linewidth]{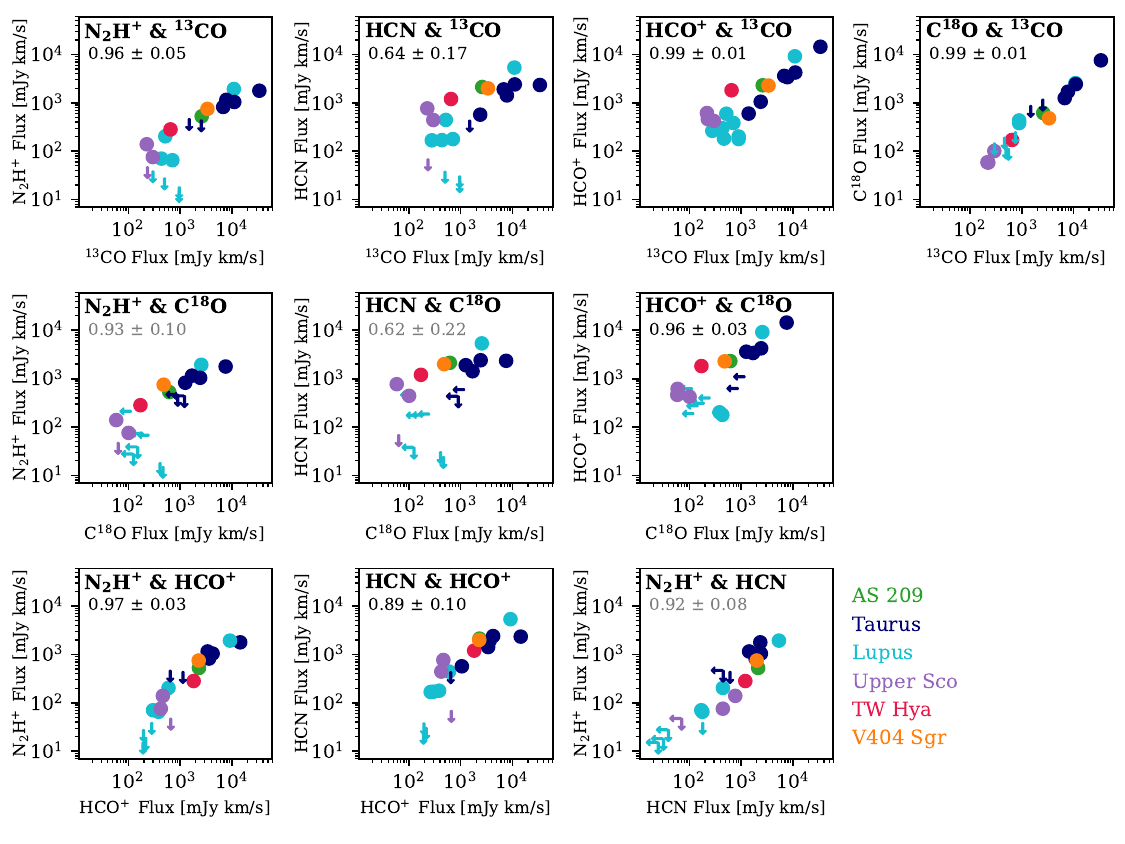}
\caption{N$_2$H$^+$, HCN, HCO$^+$, $^{13}$CO and C$^{18}$O $J$~=~3--2 observed line fluxes compared to each other. Fluxes are scaled to a distance of 160~pc. Arrows represent upper limits in either direction. Colors indicate the source or star-forming region as shown by the key in the bottom right. Computed correlation coefficients are listed in the upper left of each panel. Values listed in gray are based on only a subset of the sample, limited to sources where one or both of the fluxes values were detected. See discussion of this figure in Section~\ref{section:results5}.}
 \label{fig:FluxvsFlux}
 \end{figure*}

\begin{figure*}
\includegraphics[width=\linewidth]{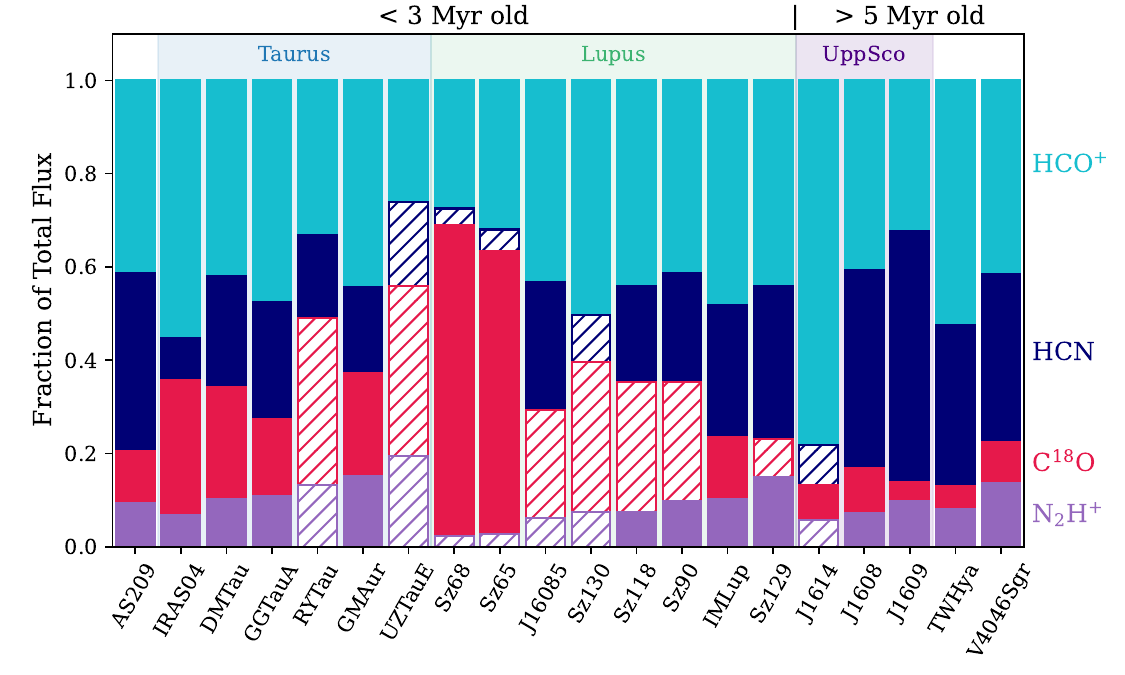}
\caption{Relative comparison of the N$_2$H$^+$, C$^{18}$O, HCN, and HCO$^+$ $J$~=~3--2 observed line fluxes for each source. Line fluxes (labeled along the right-hand side) are plotted as fractions of the total flux, where the total flux is the sum of the four line fluxes. Hatched shading indicates upper limits on fractions of non-detected lines. Sources are separated by star-forming region and age as shown by the labels along the top of the figure. See discussion of this figure in Section~\ref{section:results5}.}
 \label{fig:FluxBars}
 \end{figure*}

\subsubsection{Line Fluxes vs.~Other Stellar and Disk Properties}\label{section:results3}
In addition to the disk dust mass comparison, we also compared the molecular fluxes to other observed source properties from the literature. Through these comparisons we can investigate which source properties are influencing the chemical behavior of the disk and determine disk-integrated molecular fluxes. Data are plotted based on availability, thus not all panels contain the full 20 disks in the sample. Error bars are not included for the following parameters: M$_{star}$, L$_{star}$, M$_{acc}$, or R$_{cont,68}$. See \cite{Manara22-arxiv} for discussion on the uncertainties involved in acquiring these values. 

\cite{Anderson22} found no trends with stellar mass or luminosity. Our larger sample yields similar results (Figure~\ref{fig:FluxvsSourceProp}, columns~1--2). As shown in appendix Figure~\ref{fig:FluxvsSourceProp_A1}, normalizing the molecular line fluxes by M$_{dust}$ results in a partially decreasing trend with stellar mass and luminosity but median correlation coefficients are not more than 1--2 standard deviations above zero. This is most noticeable for HCO$^+$. The molecular fluxes show generally increasing trends with mass accretion rate, similar to their behavior with dust mass (Fig.~\ref{fig:FluxvsSourceProp}, column~3). This is expected given the known relationship between the mass accretion rates and dust masses of protoplanetary disks \citep{Manara16, Mulders17,Manara20}. However, given the smaller number of data points and some additional scatter, the linear correlation coefficients are less significantly different from zero. When normalizing the molecular line fluxes by M$_{dust}$ or the C$^{18}$O flux, there are no significant trends with M$_{acc}$ (see appendix  Figure~\ref{fig:FluxvsSourceProp_A1}).

Similar to the behavior with M$_{dust}$, molecular fluxes for all four species generally increase with the outer radius of the (sub-)millimeter dust disk (Fig.~\ref{fig:FluxvsSourceProp}, column~4). This is consistent with the idea that larger disks have more gaseous material and therefore stronger molecular line emission. However, where line emission becomes optically thick this could also be a trend in disk temperature. Perhaps larger disks have more material, pushing emitting regions of optically-thick lines to the hotter layers closer to the disk surface. The median correlation coefficients show  positive linear correlations between the molecular fluxes and R$_{cont,68}$. When normalizing the molecular line fluxes by M$_{dust}$ or the C$^{18}$O flux, there are no significant trends with R$_{cont,68}$ (see appendix Figure~\ref{fig:FluxvsSourceProp_A1}). 

\begin{figure}
\includegraphics[width=\linewidth]{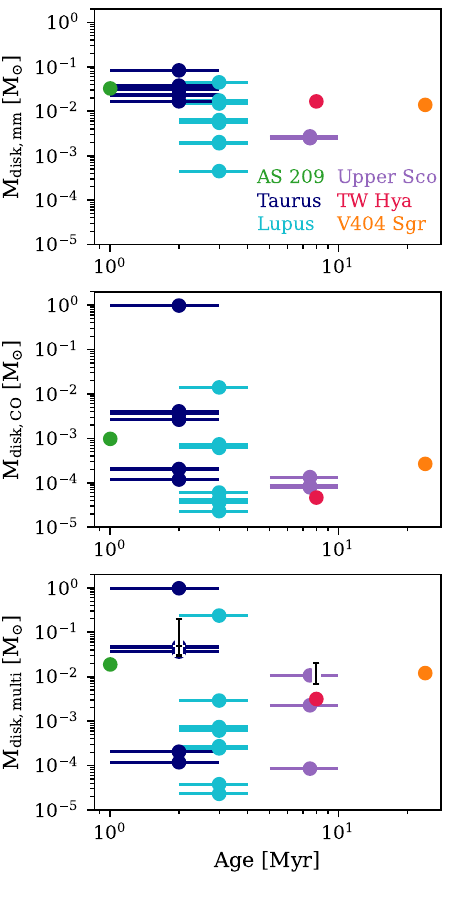}
\caption{Disk mass vs.~age for disk gas masses estimated using 100$\times$ the dust mass (top), fit functions from the models of \cite{Miotello16, Miotello17} with available observed $^{13}$CO and/or C$^{18}$O fluxes (middle), and an approximation based on the N$_2$H$^+\!$/C$^{18}$O flux ratio (bottom). The limits in the bottom panel (shown in black) are based on HD from \cite{Trapman22}. Colors indicate the source or star-forming region as indicated in the top panel. See Section~\ref{section:results6} for details on the mass estimates and Section~\ref{section:methods3} for age references.}
 \label{fig:Masses}
 \end{figure}

\subsubsection{Comparison Among Line Fluxes}\label{section:results5}
To investigate potential chemical links between molecular species, we compare pairs of molecular line fluxes from different species against each other in Figure~\ref{fig:FluxvsFlux}. \cite{Anderson22} found a tentative positive correlation among N$_2$H$^+$, HCN, and HCO$^+$ fluxes in their set of Lupus sources (8--14 in Table~\ref{table2}). These molecular fluxes were not correlated with their $^{13}$CO $J$~=~3--2 fluxes. Limited C$^{18}$O detections across their sample prevented a comparison with C$^{18}$O. In this sample of 20 sources, we find at least a tentative positive correlation in fluxes for all pairs of molecular species. The correlation coefficients computed using \texttt{linmix} (excluding any cases where both line fluxes are upper limits) show strong  positive correlations between all combinations of $^{13}$CO, C$^{18}$O, HCO$^+$, and N$_2$H$^+$. For HCN, we find strong positive correlations with HCO$^+$ and N$_2$H$^+$ and tentative positive correlations with $^{13}$CO and C$^{18}$O. Correlation coefficients are provided in Fig.~\ref{fig:FluxvsFlux}. But some of these strong correlations may break down for fainter line fluxes where constraints are lacking due to the observational bias towards brighter disks.

Correlations could be driven by physical or chemical relationships between the molecular species, or more likely, a combination of both. The strong correlation between $^{13}$CO and C$^{18}$O is unsurprising because they both track the same underlying CO abundance. In other cases, chemistry can provide explanations for positive correlations between some pairs of molecular species: CO is a precursor to the formation of HCO$^+$, N$_2$H$^+$ and HCN are both carriers of N. However, the chemical relationship between N$_2$H$^+$ and CO, that N$_2$H$^+$ is readily destroyed when CO is abundant \citep{Qi13}, would suggest they should be anticorrelated. This might be an indicator that physical disk properties, such as gas mass or temperature are playing a larger role in determining these disk-integrated molecular fluxes. The level of ionization in the disks may also be an important factor, in particular for the fluxes of molecular ions N$_2$H$^+$ and HCO$^+$.

The optical thickness of the line emission is another important factor to consider. The molecular line flux correlations generally appear less strong at lower flux values, below $\lesssim$1~Jy~km~s$^{-1}$ at a distance of 160~pc. This could be a sign of diverging pathways of chemical evolution and gas-phase chemical abundances among the fainter (typically smaller and less massive, see~Figures~\ref{fig:FluxvsMdust}--\ref{fig:FluxvsSourceProp}) disks. Alternatively, the change could represent the transition from optically thin to thick line emission, where optically thick emission is mainly tracking the gas temperature rather than the molecular column density. This would also indicate a correlation in the gas temperatures within the emitting layer of different molecular species. Isotopologue ratios (e.g., $^{13}$CO/C$^{18}$O, HCN/H$^{13}$CN) have revealed that emission from $^{13}$CO and the dominant isotologues of other molecular carriers is optically thick in a number of protoplanetary disks \citep[e.g.,][]{Schwarz16, Bergner19}. 

Figure~\ref{fig:FluxBars} compares the fluxes of all four molecular species (N$_2$H$^+$, HCN, HCO$^+$, and C$^{18}$O) against each other to give a more complete view of the variations in chemical tracers in each disk. Given the strong correlation between $^{13}$CO and C$^{18}$O, only C$^{18}$O is included to simplify the plot. HCO$^+$ makes up around 35--55\% of the total molecular flux. It is generally the largest flux contributor of the four molecular species, with the exception of the C$^{18}$O-dominant disks Sz~68 and Sz~65 and the HCN-dominant disks J1608 and J1609. The relative HCO$^+$ flux is particularly high in J1614, making up more than 80\% of the total molecular flux. The spectral shape of the HCO$^+$ flux is also an anomaly in this dataset (see Figure~\ref{fig:spectra}) suggesting special circumstances in this case. Perhaps this disk is displaying an outflow or temporary flare in HCO$^+$ emission \citep{Cleeves17,Waggoner22}. 

The relative HCN flux is highly variable, ranging from less than a few to over 40\% of the total molecular flux. It is generally higher in disks where the fraction of C$^{18}$O flux is low. The relative C$^{18}$O flux also spans a large range of values from less than a few to over 60\% of the total molecular flux. The disks with the largest C$^{18}$O fractions belong to younger star-forming regions: Taurus, Lupus, and Oph (left side of Fig.~\ref{fig:FluxBars}). But overall, the fraction of C$^{18}$O is variable across the younger disks. In contrast, C$^{18}$O fractions are all $<$10\% in the older disks (rightmost five disks in Fig.~\ref{fig:FluxBars}). Across all sources, N$_2$H$^+$ makes up less than 15\% of the total molecular flux. The lowest N$_2$H$^+$ fractions are less than a few \%, but upper limits prevent us from determining the full range and any additional patterns in N$_2$H$^+$ fractions. For at least the cases of Lupus and Upper Sco, disks from the same star-forming region display different relative fractions of molecular fluxes. This may be indicative of a range of gas chemical compositions co-existing within a single star-forming region.    

\vspace{1cm}
\subsection{Disk Gas Masses}\label{section:results6}
Total disk gas mass is an essential parameter regulating planet formation mechanisms and the number of planets that can form in a given system. Disk masses are dominated by H$_2$ gas, which is unobservable at cold temperatures throughout the bulk of the disk. As such, indirect tracers, most often sub-mm dust and CO isotopologue emission, are traditionally used to provide disk gas mass estimates. While both are readily observable and useful for a first look, the accuracy of the resulting gas mass estimates ultimately depends on the accuracy of the gas/dust or CO/H$_2$ conversion factor. The dust may evolve separately from the gas over time, resulting in uncertainty in the gas/dust ratio. CO/H$_2$ can also vary spatially and temporally within the disk and among different systems \citep[e.g.,][]{Schwarz16, Zhang19, Zhang21-MAPS}. Measuring CO/H$_2$ values in protoplanetary disks requires at least one additional measured quantity beyond CO isotopologue emission, preferably one that is tied to the H$_2$ content of the gas. HD measurements enabled by the Herschel Space Observatory \citep{Bergin13,McClure16} revealed CO/H$_2$ up to 100$\times$ below the typically assumed interstellar value of $\sim$10$^{-4}$ in a few protoplanetary disks. Since the end of Herschel's mission, there have not been any observatories equipped for the necessary HD measurements. Meanwhile, developments have been made in using N$_2$H$^+$ in combination with CO to provide constraints on both the CO/H$_2$ and H$_2$ mass \citep{Anderson19,Anderson22,Trapman22}. N$_2$H$^+$ is readily observable by ground-based radio observatories including ALMA and the SMA. 

We use our data to compare disk gas mass estimates based on measurements of (sub-)mm dust, CO gas, and from combinations of molecular gas tracers. In Figure~\ref{fig:Masses}, we plot an initial look at how disk gas masses vary with age across our selected sample. The top panel shows gas masses as equal to 100$\times$ M$_{dust}$. Dust masses are computed using (sub-)millimeter continuum emission as described in Section~\ref{section:methods3}. More than half of this sample lies above 0.01~M$_{\odot}$, with the remaining sources extending to values 1--2 orders of magnitude lower. The Lupus sources represent values across the full range, whereas Taurus sources and Upper Sco sources are each grouped around a narrower range of disk dust mass values. Previous comparisons across star-forming regions find decreasing median dust masses with age \citep[e.g.,][]{Ansdell17,Cieza19,Villenave21}. The sources in this sample were not selected to be representative of the population in their respective star-forming regions. Selection is often biased towards brighter sources, therefore this comparison is not representative of typical sources across the star-forming regions included in this work.

The second panel in Fig.~\ref{fig:Masses} shows gas masses based on CO isotopologue observations. These masses were computed using the fits functions provided by \cite{Miotello16,Miotello17} from their large grid of models. Masses were based on C$^{18}$O fluxes from either the $J$~=~3--2 or $J$~=~2--1 transitions using the appropriate functions for each from \cite{Miotello16,Miotello17}. Where fluxes for both transitions are available, the one with a higher signal-to-noise ratio was used. For sources where C$^{18}$O has not been detected, $^{13}$CO fluxes and appropriate functions are used instead. With the exception of the young source IRAS~04302+2247, CO-based gas masses are lower than the dust-based values by roughly 1--2 orders of magnitude.  While IRAS~04302+2247 is part of the Taurus star-forming region and that age is used here, it may represent an earlier evolutionary stage as it is classified as a Class~I source \citep{Lucas97, Garufi21}. The distinction between gas masses derived from (sub-)millimeter dust and those derived from CO isotopologues is well documented in large population ALMA studies of multiple star-forming regions \citep[e.g.,][]{Ansdell16,Miotello17,Long17}. As discussed in these previous works, the differences between gas mass estimates derived from these two methods could be the result of gas/dust ratios and/or CO/H$_2$ abundances decreasing over time as the disks evolve. 

\cite{Anderson19,Anderson22} and \cite{Trapman22} have shown that by using a combination of N$_2$H$^+$ and CO observations, we can place constraints on both the total disk gas mass and CO/H$_2$. In this way, we can distinguish between the effects of depleted gas/dust ratios vs.~depleted CO/H$_2$. The extensive modeling analysis needed to constrain CO/H$_2$ and H$_2$ masses for this sample of disks is beyond the scope of this work. But to give a rough idea of how this might affect our inferred gas masses, we perform a simple scaling of the CO-based  total H$_2$ gas masses from the middle panel of Fig.~\ref{fig:Masses}. These values assume a CO/H$_2$ abundance of $\sim$10$^{-4}$. We increase these values by a factor corresponding to the CO/H$_2$ abundances derived from the observed N$_2$H$^+$/C$^{18}$O flux ratios for each source. For CO/H$_2$, we used the modeled relation between the N$_2$H$^+$ 3--2 / C$^{18}$O 2--1 flux ratio and the CO/H$_2$ abundance for the TW~Hya model with a cosmic ray ionization rate of 10$^{-18}$~s$^{-1}$ from \cite{Trapman22}. In reality, this relation will depend on the individual physical/chemical structure of each disk \citep{Trapman22} and further exploration of how much this relation varies over a wider parameter space is needed. 

Based on this approximation, the gas masses for a subset of our sample were increased by factors of $\sim$5--135$\times$ relative to the CO-based values. Some disks have N$_2$H$^+$\!/C$^{18}$O flux ratios that are consistent with a CO/H$_2$ of 10$^{-4}$, including disks where only an upper limit on this flux ratio is available, so their gas mass estimates were not changed from the CO-based values. The gas mass value for IRAS~04302+2247 was also not increased given its already high value. The final result is a larger spread in the range of gas masses for each age group (Fig.~\ref{fig:Masses}, bottom panel). The additional constraints on CO/H$_2$ increase the gas mass estimates for some disks to values that are closer to the dust-based masses (Fig.~\ref{fig:Masses}, top panel). Mass estimates based on HD emission \citep{Bergin13, McClure16, Trapman22} also suggest that the CO-based values are underestimating the total amount of H$_2$ gas present. Our conclusions are limited by the size of and selection biases present in the current sample. Ultimately, more data are needed to explore the evolution of disk gas masses, especially data from older sources (ages $>$5~Myr).  

\section{Discussion}\label{section:discussion}
\subsection{Comparison within our sample}
This dataset includes 20 disks spread across various birth environments and ages. Within this sample, we find examples of disks with very different disk integrated molecular flux ratios, even for disks residing in the same star-forming region. In Figures~\ref{fig:FluxvsMdust}--\ref{fig:FluxvsSourceProp} color is used to distinguish sources from different star-forming regions or moving groups. In this case, any apparent disparities among disks from different star-forming regions are likely caused by differences in the sample selection criteria used for the different sets of observations. As a result, the selected Taurus and individual sources (AS~209, TW~Hya, and V4046~Sgr) tend to have larger and brighter disks relative to those from Lupus and Upper Sco. When normalized by M$_{dust}$ or C$^{18}$O, molecular fluxes fall within generally the same range within 1-2 orders of magnitude regardless of star-forming region---with a few exceptions. There are two sources from Lupus (Sz~65 and Sz~68, both appearing as upper limits in Figures~\ref{fig:FluxvsSourceProp_A1}--\ref{fig:FluxvsSourceProp_A2}) for which constraints on the normalized N$_2$H$^+$ and HCN values fall well below the rest of the sample. This could be an indication that these two sources have a distinct disk gas chemical composition from the others. 

 For the comparisons made in Figures~\ref{fig:FluxvsMdust}--\ref{fig:FluxvsFlux}, we see no obvious distinctions between the younger sources (AS~209, Taurus, and Lupus disks; 1--3~Myr old) and the older sources (Upper Sco, TW~Hya, and V4046~Sgr disks; $>$5~Myr old). Although, for at least 4/5 of the older sources, their N$_2$H$^+$, HCN, and HCO$^+$ fluxes normalized by M$_{dust}$ or C$^{18}$O are among the highest values in the sample. There are hints of more uniformity among these disks (Fig.~\ref{fig:FluxBars}). The fifth source, J1614, stands out by having a relatively large amount of HCO$^+$ emission with an unusual line shape. However, even when excluding HCO$^+$, J1614 still differs from the other $>$5~Myr-old sources due to its low HCN/C$^{18}$O flux ratio. 
 
 It should be noted that different age ranges are not uniformly sampled in this work. The older populations may suffer the greatest selection bias, often towards brighter objects that still have molecular gas. Overall, aged populations are more poorly sampled in multi-molecular observations. Furthermore, this comparison may be muddled by uncertainty in individual source ages and potential age variations within star-forming regions \citep[e.g.,][]{Krolikowski21}. A more uniform unbiased survey of disks across different age brackets is needed to fully investigate trends in molecular fluxes with disk age. 

Sample bias and incompleteness largely affect this current analysis. The selected sources tend to be among the largest and brightest in their respective star-forming regions. The star-forming regions included in this comparison are also not equally sampled, with fewer sources particularly for the older populations. Furthermore, because the sample is compiled from different observing programs, the sensitivity is not the same for all sources or molecular lines. Additional uncertainty is introduced when trying to make comparisons using different transitions of CO isotopologues (see Section~\ref{section:methods3}). Current ALMA Large Programs in progress, namely AGE-PRO: The ALMA Survey of Gas Evolution in PROtoplanetary Disks and The ALMA Disk-Exoplanet C/Onnection (ALMA DECO), will provide a more uniform sampling and analysis of multiple molecular lines across different star-forming regions. 

Within our sample are four systems that contain multiple stars. Given the current dataset, there are no obvious flux trends that set these systems apart from the single star systems. Although it should be noted that the datasets of stellar and disk properties are not complete for our sample, especially for the stellar multiple systems. In addition, using spatially resolved observations that can distinguish between different disk components in these complex systems (e.g., disks around individual stars vs.~disks around binaries) may be necessary to make more meaningful comparisons. 

There are many relevant parameters that could affect the observed molecular line fluxes that have not been explored in this work. This is largely due to limitations in the data readily available for this combined sample. Additional stellar properties such as UV and X-ray spectra, stellar activity, and magnetic field strength could be explored for connections in the future. Fluxes will also depend on the individual density and temperature distribution of each disk. Determining whether these physical conditions are relatively consistent across populations (or sub-groups within populations) or unique to each individual source is an important step in understanding disk evolution and planet-forming environments.  

\subsection{Comparison with previous molecular disk surveys}
Previous works have explored surveys of disk-integrated molecular line fluxes for different samples of disks and molecular species. \cite{Pegues23} find that molecular line fluxes generally increase with continuum fluxes across their sample of T~Tauri and Herbig disks (stellar masses 0.15--2.5~M$_{\odot}$), consistent with our trends in line flux with M$_{dust}$. Our results agree with \cite{Pegues21} in finding no correlation between the flux ratio of HCN/C$^{18}$O and M$_{star}$ (Figure~\ref{fig:FluxvsSourceProp_A2}). \cite{Pegues23} also show that HCO$^+$/C$^{18}$O flux ratios (when both species are detected) are relatively flat with continuum flux and stellar luminosity for their T~Tauri sample, similar to our findings.

While we find positive correlations between the disk-integrated line fluxes of all our molecular species pairs, this is not always the case. \cite{vanTerwisga19} found that CN fluxes do not strongly correlate with $^{13}$CO in disks from the Lupus star-forming region. Furthermore, the CN fluxes do not strongly correlate with the sub-millimeter continuum fluxes. However, for a subset of the sample \citep{Miotello19} show a strong correlation between CN and C$_2$H fluxes and between both species and the stellar luminosity. The abundances of CN and C$_2$H are largely affected by photochemistry, which might explain their correlation.   

With a diverse set of 14 disks across a range of ages, stellar masses, and stellar luminosities (including some overlap with our sample), \cite{Bergner19} find no correlation between disk-integrated HCN and C$^{18}$O fluxes when both are normalized to the continuum flux. Here we find only a tentative trend between the HCN and C$^{18}$O fluxes relative to strong positive correlations between other molecular species. In contrast, \cite{Pegues21} report a significant positive correlation between HCN and C$^{18}$O fluxes, although the correlation is not as strong as those between other molecular species pairs (C$_2$H and HCN, H$_2$CO and HCN). In addition, they find a positive correlation between C$^{18}$O and M$_{star}$, whereas we find no correlation. The sample of \cite{Pegues21} includes disks around five low mass stars (0.14--0.23~M$_{\odot}$) alongside solar-type T~Tauri and Herbig Ae disks. In comparison, our sample is more concentrated towards stellar masses of $\sim$0.5--1~M$_{\odot}$ and does not include such low mass stars ($<$0.25~M$_{\odot}$). Current trends are determined with limited numbers of disks and line detections, more sensitive observations for larger and more diverse samples of disks are ultimately needed to explore these discrepancies as well as to verify existing trends.

\subsection{Comparison with physical/chemical disk models}
Molecular abundances are sensitive to various physical and chemical properties of protoplanetary disks \citep{Oberg23}. Molecular emission is further sensitive to excitation conditions, predominately how the abundance distribution aligns with the disk temperature and density structures. Physical/chemical disk models have explored how the molecular fluxes of some of our observed species depend on various stellar and disk parameters. \cite{Miotello16} show that disk-integrated $^{13}$CO and C$^{18}$O fluxes are sensitive to the disk mass and outer radial extent. The modeling of \cite{Boyden23} shows HCO$^+$ $J$~=~4--3 emission increases with the outer disk radius, but decreases with increasing stellar mass and disk dust mass. \cite{Fedele20} find HCN $J$~=~4--3 emission is insensitive to changes in stellar and physical disk parameters, including the total stellar luminosity, but sensitive to elemental ratios of C/O (particularly as C/H is reduced) and N/H. But note that the $J$~=~3--2 HCO$^+$ and HCN transitions observed here have lower upper state energies and likely trace a colder and therefore deeper vertical layer of the disk. \cite{Anderson22} show that $^{13}$CO, HCO$^+$, HCN, and N$_2$H$^+$ $J$~=~3--2 emission generally increase together for increasing gas/dust ratios, but diverge with volatile depletion (C/H and O/H). As C and O are removed, $^{13}$CO and HCO$^+$ emission decreases, while HCN and N$_2$H$^+$ emission increases. 

While our observed trends agree with some of the model results (e.g., increasing line emission with disk radius), direct comparison with existing model grids is difficult. All model results depend on the assumptions made about the interplay of different disk parameters, for example: the relationship between gas and dust components of the disk and between disk mass and outer radius. The assumed relationships may not be reflective of our observed sample of disks. While beyond the scope of this work, a model grid exploring N$_2$H$^+$, HCN, HCO$^+$, C$^{18}$O, and $^{13}$CO $J$~=~3--2 emission across the parameter space relevant to the star and disk properties of this observed sample would aid in the interpretation of the trends we find.

\vspace{0.5cm}
\section{Conclusion}\label{section:conc}
In this work, we present observations of N$_2$H$^+$, HCN, HCO$^+$, and CO isotopologues in 4 disks from the young Taurus star-forming region and 3 disks from the older Upper Scorpius star-forming region. Using data from the literature, we create a sample of 20 disks from multiple environments, including five sources at $\gtrsim$5~Myr old. We compare the observed molecular line fluxes with stellar and disk properties from the literature to look for connections indicative of underlying physical and/or chemical disk properties. Overall, we find molecular line fluxes generally increase with (sub-)millimeter disk dust masses and outer radii. But there is a great deal of scatter within these relationships. Strong positive correlations exist among the line fluxes of different molecular species. However, scatter in these relations indicate that even for disks in a single star-forming region---presumably of similar age and birth environment---there are differences in physical and/or chemical properties that affect the relative molecular line fluxes of N$_2$H$^+$, HCN, HCO$^+$, and C$^{18}$O. 

This sample has been compiled from numerous separate observing programs. The lack of uniformity in source selection and observing parameters certainly affect our current results. Future observing campaigns aimed at producing more uniform surveys of multiple molecular species in large populations of protoplanetary disks are needed to provide a more complete analysis of the tentative relationships in observed disk properties investigated here. 

\software {Astropy \citep{astropy:2013, astropy:2018}, CASA \citep{2007ASPC..376..127M}, linmix (\url{https://github.com/jmeyers314/linmix}), Matplotlib \citep{matplotlib}, Numpy \citep{numpy}}

\acknowledgments D.E.A. acknowledges support from the Virginia Initiative on Cosmic Origins (VICO) Postdoctoral and Carnegie Postdoctoral Fellowships. L.I.C. gratefully acknowledges support from the David and Lucille Packard Foundation, the Virginia Space Grant Consortium, Johnson \& Johnson's WiSTEM2D Award, and NSF AAG grant number AST-1910106. This paper makes use of the following ALMA data: ADS/JAO.ALMA\#2015.1.01199.S, ADS/JAO.ALMA\#2018.1.01623.S, and \\ADS/JAO.ALMA\#2019.1.01135.S. ALMA is a partnership of ESO (representing its member states), NSF (USA) and NINS (Japan), together with NRC (Canada), MOST and ASIAA (Taiwan), and KASI (Republic of Korea), in cooperation with the Republic of Chile. The Joint ALMA Observatory is operated by ESO, AUI/NRAO and NAOJ. This paper also makes use of data from Submillimeter Array project 2018B-S046. The Submillimeter Array is a joint project between the Smithsonian Astrophysical Observatory and the Academia Sinica Institute of Astronomy and Astrophysics and is funded by the Smithsonian Institution and the Academia Sinica. This work has made use of data from the European Space Agency (ESA) mission
{\it Gaia} (\url{https://www.cosmos.esa.int/gaia}), processed by the {\it Gaia}
Data Processing and Analysis Consortium (DPAC,
\url{https://www.cosmos.esa.int/web/gaia/dpac/consortium}). Funding for the DPAC
has been provided by national institutions, in particular the institutions
participating in the {\it Gaia} Multilateral Agreement.

\bibliography{bibliography2021}

\appendix
Here we present the spectral window settings for the ALMA observations in Table~\ref{tableA1}, the moment zero maps of integrated flux from the observations described in Sections~\ref{section:methods1a}--\ref{section:methods1b} in Figure~\ref{fig:maps}, and the main CO isotopologue emission for observed Taurus sources in Figure~\ref{fig:CO_spectra}. Additional figures are included to show further comparisons between observed line fluxes normalized by M$_{dust}$ (Figure~\ref{fig:FluxvsSourceProp_A1}) and C$^{18}$O line fluxes (Figure~\ref{fig:FluxvsSourceProp_A2}) and source properties. A summary of the median correlation coefficients from the \texttt{linmix} analyses of parameter pairs are also presented in Figure~\ref{fig:stats}.

\renewcommand{\thefigure}{A\arabic{figure}}
\setcounter{figure}{0}

\renewcommand{\thetable}{A\arabic{table}}
\setcounter{table}{0}

\begin{deluxetable*}{lclllrcc}[h!]
\tablewidth{0pt}
\setlength{\tabcolsep}{4pt}
\tablecolumns{8}
\tablecaption{\em{Observational Settings}\label{tableA1}}
\tablehead{ \multicolumn{1}{l}{Dataset} & \multicolumn{1}{c}{ALMA} & \multicolumn{1}{l}{Target} & \multicolumn{3}{c}{Spectral Windows} & \multicolumn{1}{c}{Beam Size} & \multicolumn{1}{c}{Int.$^a$}\\
 & \multicolumn{1}{c}{Band} & & \multicolumn{1}{l}{Width} & \multicolumn{1}{l}{Center} & \multicolumn{1}{l}{Channel}  & & \multicolumn{1}{c}{(min)} \\
 & & & \multicolumn{1}{l}{(GHz)} & \multicolumn{1}{l}{(rest, GHz)} & \multicolumn{1}{l}{width (kHz)}  & &  }
\startdata
2015.1.01199.S & 7 & Spectral lines & 0.117 & 279.51176 (N$_2$H$^+$ J~=~3--2) & 122.070 & $\sim$0.6\arcsec$\times$0.5\arcsec & 107 \\
Sources:~J1608, J1609 & &  &  & 281.52693 &  &  &  \\
  &  &  &  & 282.38109 &  &  &  \\
   &  &  &  & 282.92001 &  &  &  \\
      &  &  &  & 293.91209 &  &  &  \\
& & Continuum & 1.875 & 293.90000  & 976.562 & &\\
2018.1.01623.S & 6 & Spectral lines & 0.234 & 255.47939 & 244.141 & $\sim$0.4\arcsec$\times$0.4\arcsec & 23 \\
 Sources:~J1608, J1609, & &  &  & 265.88643 (HCN J~=~3--2) &  &  &  \\
J1614 &  &  &  & 267.55763 (HCO$^+$ J~=~3--2) &  &  & \\
& & Continuum & 1.875 & 251.80000  & 976.562 & & \\
& &  & & 269.50000  & 15625.000 & &\\
& 7 & Spectral lines & 0.469 & 279.51176 (N$_2$H$^+$ J~=~3--2)  & 244.141 & $\sim$0.4\arcsec$\times$0.4\arcsec & 36 \\
 &  & & 0.117 & 288.14386 &  & & \\ 
 & &  & & 289.20907  & & &\\
 & &  & & 289.64491  & & &\\
 & & Continuum & 1.875 & 278.20000   & 15625.000 & & \\
  & &  & & 291.05000 & & &\\
 & 7 & Spectral lines & 0.234 & 329.33055 (C$^{18}$O J~=~3--2)  & 244.141 & $\sim$0.8\arcsec$\times$0.6\arcsec & 57 \\
 & &  &  &  330.587967 ($^{13}$CO J~=~3--2)  &  &  &  \\
 & & & 0.469 & 340.24777 (CN N~=~3--2)  &  & &  \\
  & & Continuum & 1.875 & 329.33055   & 976.562 & & \\
    & &  & & 341.50000 & & &\\
\enddata
\tablecomments{$^a$Approximate on-source time per source.}
\end{deluxetable*}

 \begin{figure*}
 \includegraphics[width=\linewidth]{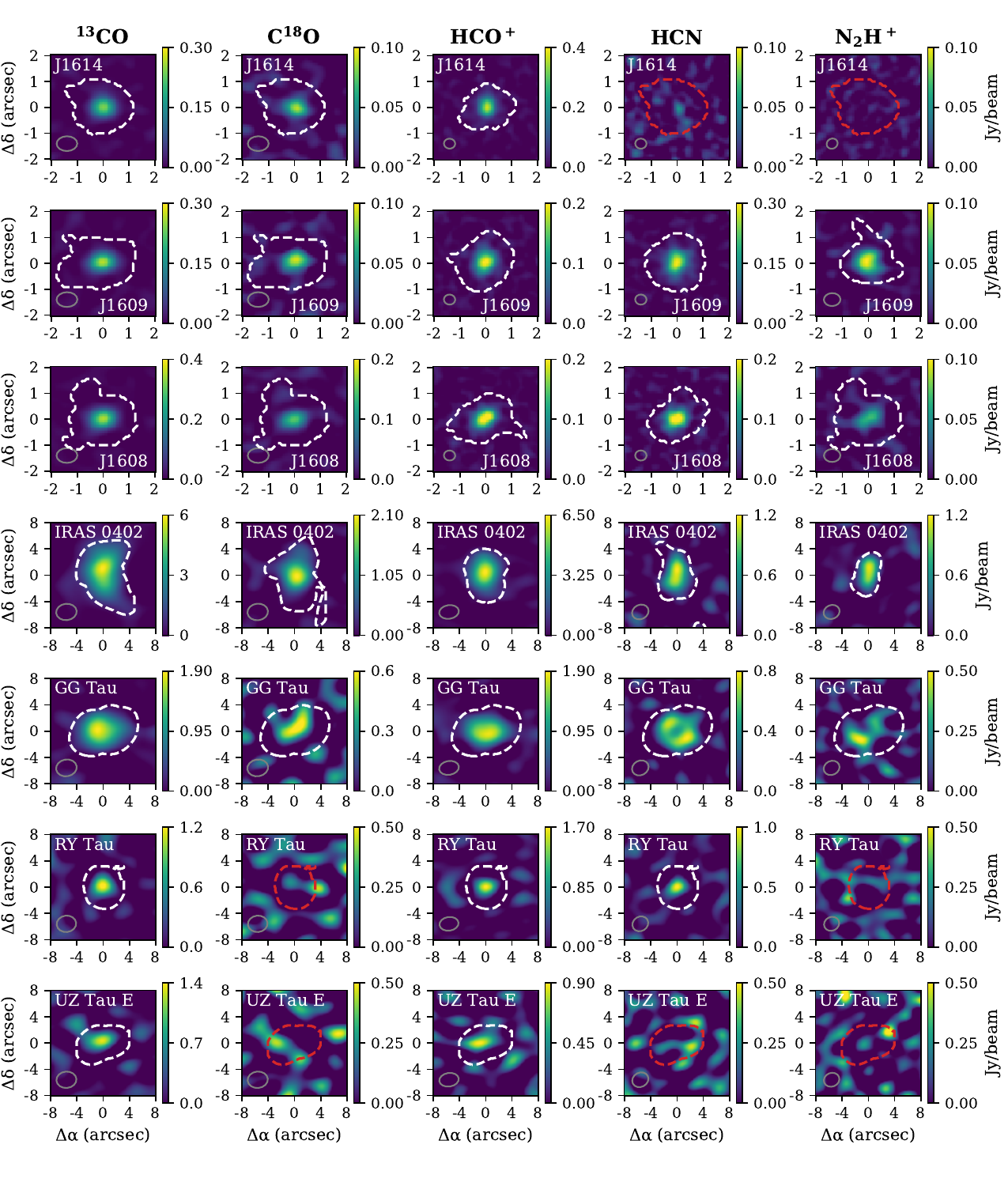} 
 \caption{Moment 0 maps of integrated flux for $^{13}$CO, C$^{18}$O ($J$~=~3--2 for the top three rows, 2--1 for the bottom four), HCO$^+$, HCN, and N$_2$H$^+$ ($J$~=~3--2) line emission from the observed disks described in Sections~\ref{section:methods1a}--\ref{section:methods1b}. Fluxes are integrated over the velocity range of -5--15~km~s$^{-1}$ for Upper Sco sources (-15--20~km~s$^{-1}$ for HCO$^+$ in J1614), 4--9~km~s$^{-1}$ for GG~Tau~A, 2--8~km~s$^{-1}$ for IRAS~04302+2247, 2--10~km~s$^{-1}$ for RY~Tau, and 2--13~km~s$^{-1}$ for UZ~Tau~E. Velocity ranges are based on observed line emission and ranges are inclusive for 1~km~s$^{-1}$ channels. Overplotted contours indicate the summed empirical masks used to encapsulate regions containing potential line emission. Red contours indicate non-detections. \label{fig:maps}}
 \end{figure*}

 \begin{figure*}
 \includegraphics[width=\linewidth]{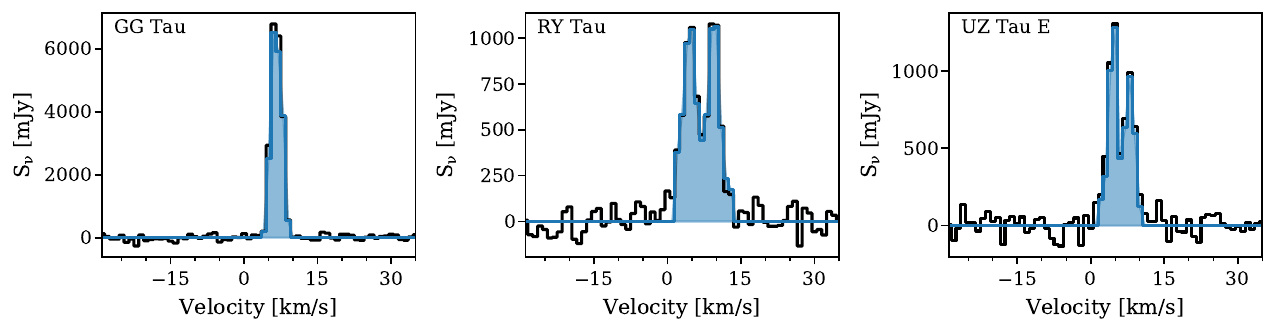}
 \caption{Spectra showing CO $J$~=~2--1 line emission from the observed disks described in Section~\ref{section:methods1b}. Black curves show the integrated flux from within the channel-summed empirical mask, whereas the integrated fluxes within the channel-by-channel empirical masks are shown in blue. This CO emission was used to generate the empirical masks for fainter lines in these sources. Integrated CO fluxes (computed as described in Section~\ref{section:methods2}) are 19547$\pm$1985~mJy~km~s$^{-1}$ for GG~Tau~A, 7679$\pm$794~mJy~km~s$^{-1}$ for RY~Tau, and 5523$\pm$584~mJy~km~s$^{-1}$ for UZ~Tau~E. The CO spectrum for IRAS~04302+2247 shows signs of absorption and is not used in this analysis.}
 \label{fig:CO_spectra}
 \end{figure*}

 \begin{figure*}
 \includegraphics[width=\linewidth]{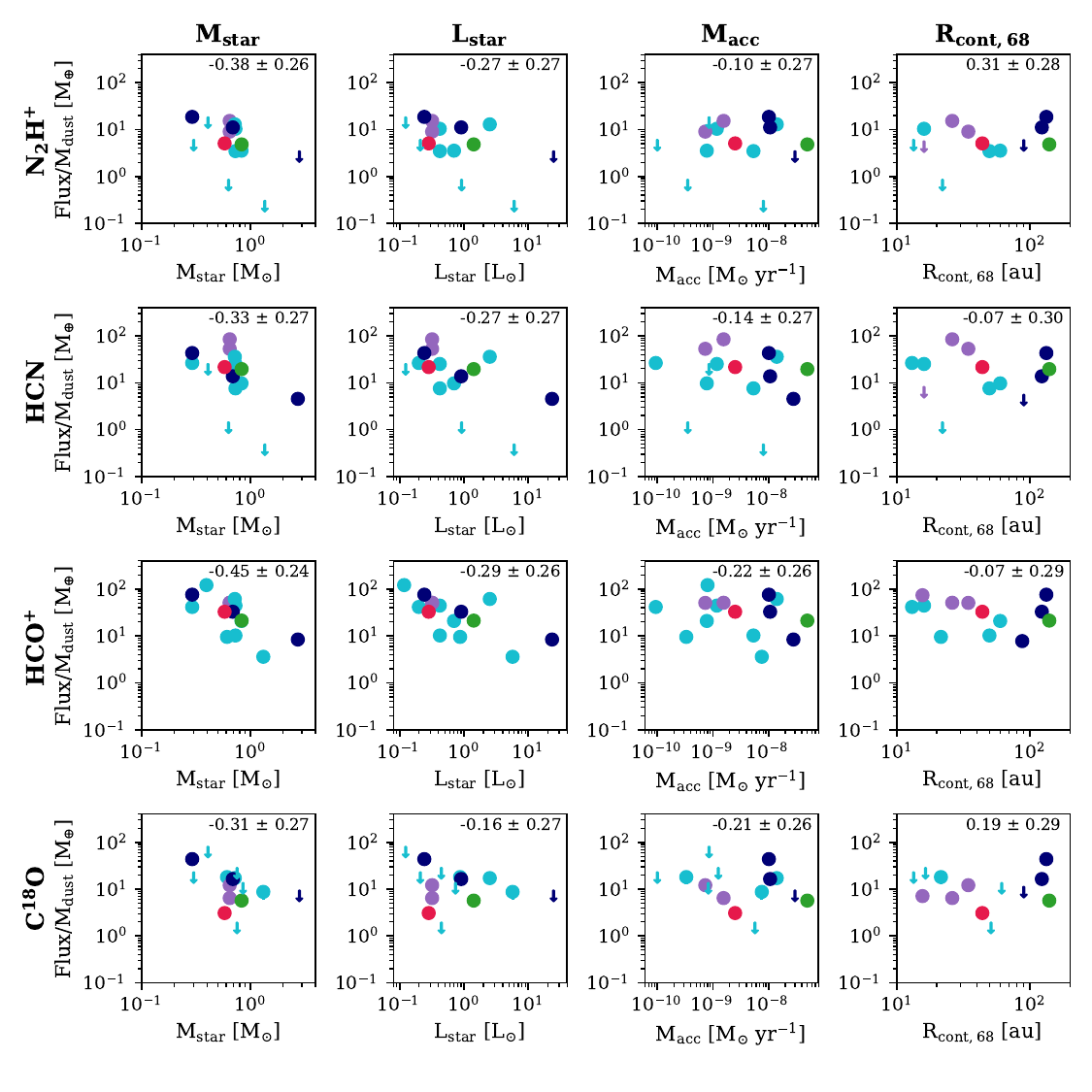}
 \caption{Variation of Figure~\ref{fig:FluxvsSourceProp} showing line fluxes normalized by disk dust masses. Normalized N$_2$H$^+$, HCN, HCO$^+$, and C$^{18}$O $J$~=~3--2 observed line fluxes relative to disk dust masses (rows) compared to stellar masses (first column), stellar luminosities (second column), mass accretion rates (third column),  and outer disk radius seen in (sub-)millimeter dust (fourth column). Fluxes are scaled to a distance of 160~pc. Arrows represent upper limits. Colors indicate the source or star-forming region as indicated in Figure~\ref{fig:FluxvsMdust}. Only a subset of the sample is shown per panel as literature data were not equally available for all sources (see Table~\ref{table1}). Computed correlation coefficients are provided in the upper right of each panel.}
 \label{fig:FluxvsSourceProp_A1}
 \end{figure*}

 \begin{figure*}
 \includegraphics[width=\linewidth]{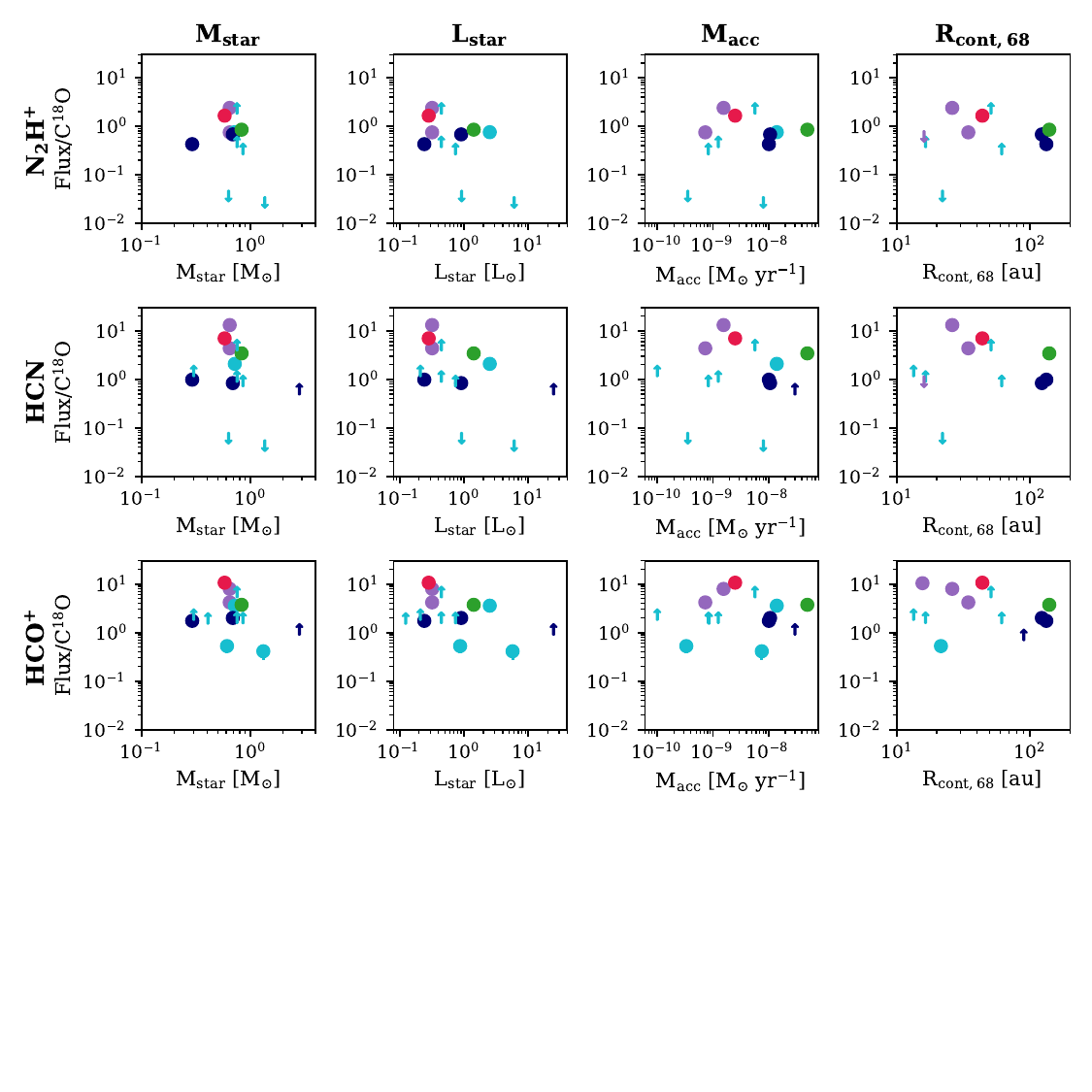}
 \caption{Variation of Figure~\ref{fig:FluxvsSourceProp} showing line fluxes normalized by the C$^{18}$O line flux. Normalized N$_2$H$^+$, HCN, and HCO$^+$ $J$~=~3--2 observed line fluxes relative to C$^{18}$O (rows) compared to stellar masses (first column), stellar luminosities (second column), mass accretion rates (third column),  and outer disk radius seen in (sub-)millimeter dust (fourth column). Fluxes are scaled to a distance of 160~pc. Arrows represent upper or lower limits. Colors indicate the source or star-forming region as indicated in Figure~\ref{fig:FluxvsMdust}. Only a subset of the sample is shown per panel as literature data were not equally available for all sources (see Table~\ref{table1}). Correlation coefficients are not provided because there are no upper or lower bounds on flux ratios.}
 \label{fig:FluxvsSourceProp_A2}
 \end{figure*}

 \begin{figure}
 \includegraphics[width=\linewidth]{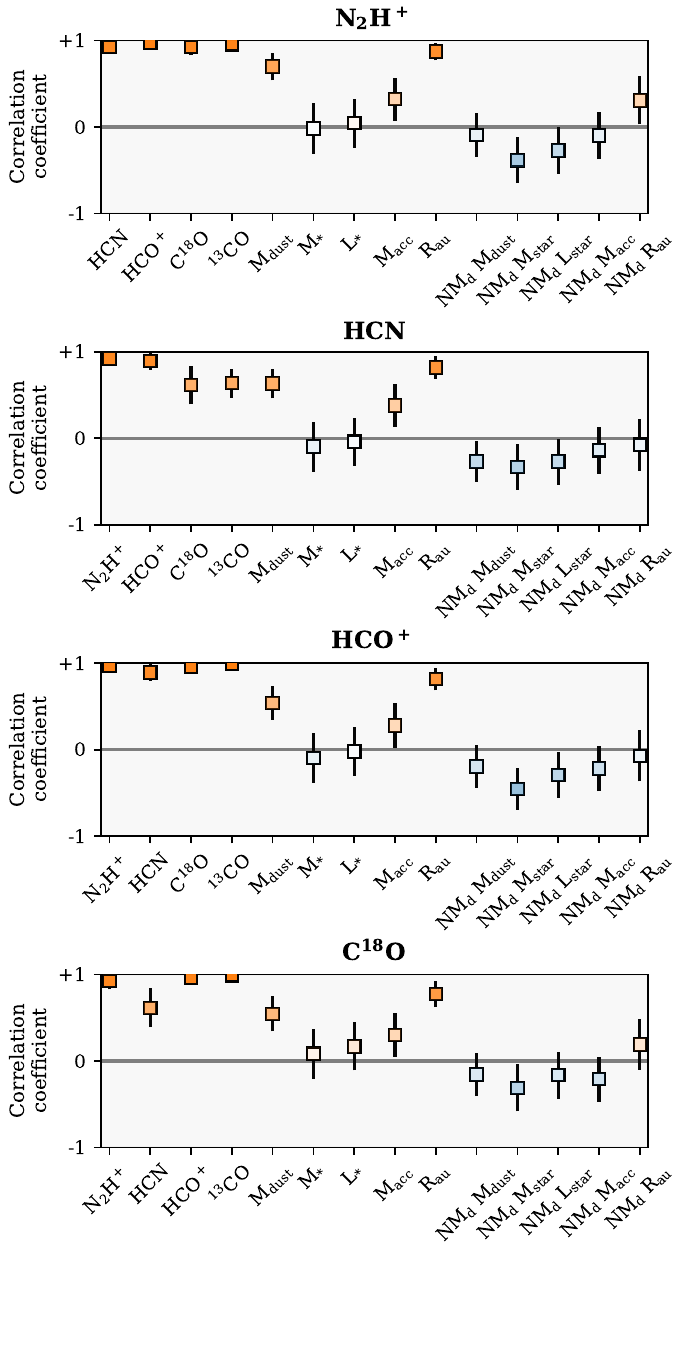}
 \caption{Median linear correlation coefficients for each source parameter (listed on the x-axis) with observed N$_2$H$^+$, HCN, HCO$^+$, and C$^{18}$O $J$~=~3--2 fluxes. NM$_d$ denotes the molecular line fluxes normalized by M$_{dust}$ (see Figures~\ref{fig:FluxvsMdust} and \ref{fig:FluxvsSourceProp_A1}). Error bars represent one standard deviation in the linear correlation coefficients. Positive correlations (with coefficients from 0 to 1) are shown in orange and anticorrelations (0 to -1) in blue. Darker colors represent stronger (anti)correlations based on the median value. Values close to zero indicate no linear correlation between the two parameters.}
 \label{fig:stats}
 \end{figure}

\end{document}